\documentstyle[12pt,aaspp4]{article}

\lefthead{Rey et al.}
\righthead{GALEX UV Photometry of M31 Globular Clusters}

\begin{document}
\title{\large\textbf{ {\sl GALEX} Ultraviolet Photometry of Globular Clusters in M31:  Three Year Results and a  Catalog}\normalsize\textnormal{}}

\author{Soo-Chang Rey\altaffilmark{1,2,3}, R. Michael Rich\altaffilmark{4}, Sangmo T. Sohn\altaffilmark{2,5}, 
Suk-Jin Yoon\altaffilmark{2}, Chul Chung\altaffilmark{2}, Sukyoung K. Yi\altaffilmark{2}, Young-Wook Lee\altaffilmark{2}, 
Jaehyon Rhee\altaffilmark{2,3}, Luciana Bianchi\altaffilmark{6}, Barry F. Madore\altaffilmark{7}, 
Kyungsook Lee\altaffilmark{1}, Tom A. Barlow\altaffilmark{3}, Karl Forster\altaffilmark{3}, Peter G. Friedman\altaffilmark{3},
D. Christopher Martin\altaffilmark{3}, Patrick Morrissey\altaffilmark{3}, Susan G. Neff\altaffilmark{8},
David Schiminovich\altaffilmark{9}, Mark Seibert\altaffilmark{3}, Todd Small\altaffilmark{3},
Ted K. Wyder\altaffilmark{3}, 
Jose Donas\altaffilmark{10}, Timothy M. Heckman\altaffilmark{11}, Bruno Milliard\altaffilmark{10},
Alex S. Szalay\altaffilmark{11}, Barry Y. Welsh\altaffilmark{12}
}

\altaffiltext{1}{Department of Astronomy and Space Science, 
Chungnam National University, Daejeon 305-764, Korea; screy@cnu.ac.kr}
\altaffiltext{2}{Center for Space Astrophysics, Yonsei University, Seoul
120-749, Korea}
\altaffiltext{3}{California Institute of Technology, MC 405-47, 1200 East
California Boulevard, Pasadena, CA 91125}
\altaffiltext{4}{Department of Physics and Astronomy, University of
California, Los Angeles, CA 90095}
\altaffiltext{5}{Korea Astronomy and Space Science Institute, 61-1 Hwaam-dong,
Yuseong-gu, Daejeon 305-348, Korea}
\altaffiltext{6}{Center for Astrophysical Sciences, The Johns Hopkins
University, 3400 N. Charles St., Baltimore, MD 21218}
\altaffiltext{7}{Observatories of the Carnegie Institution of Washington,
813 Santa Barbara St., Pasadena, CA 91101}
\altaffiltext{8}{Laboratory for Astronomy and Solar Physics, NASA Goddard
Space Flight Center, Greenbelt, MD 20771}
\altaffiltext{9} {Department of Astronomy, Columbia University, New York, NY 10027}
\altaffiltext{10}{Laboratoire d'Astrophysique de Marseille, BP 8, Traverse
du Siphon, 13376 Marseille Cedex 12, France}
\altaffiltext{11}{Department of Physics and Astronomy, The Johns Hopkins
University, Homewood Campus, Baltimore, MD 21218}
\altaffiltext{12}{Space Sciences Laboratory, University of California at
Berkeley, 601 Campbell Hall, Berkeley, CA 94720}

\begin{abstract}

We present ultraviolet (UV) photometry of M31 globular clusters (GCs) 
found in 23 {\sl Galaxy Evolution Explorer} ({\sl GALEX}) 
images covering the entirety of M31.  We detect 485 and 273 GCs 
(and GC candidates) in the near-ultraviolet (NUV; $\lambda_{eff}$ 
= 2267 \AA, $\Delta\lambda$ = 732 \AA) and far-ultraviolet 
(FUV; $\lambda_{eff}$ = 1516 \AA, $\Delta\lambda$ = 268 \AA), 
respectively.  Our UV catalog has been complemented with existing 
optical and near-infrared photometry.
The UV properties of GCs have been analyzed using various combinations 
of UV$-$optical and optical$-$optical colors considered to 
be tracers of ages and metallicities for simple stellar 
populations.  Comparing M31 data with those of Galactic GCs in the 
UV with the aid of population models that fully take into account 
detailed systematic variation of horizontal-branch (HB) morphology 
with age and metallicity, we find that the age ranges of old GCs in 
M31 and the Galactic halo are similar.
Three metal-rich ([Fe/H]$>-1$) GCs in M31 produce significant FUV flux 
making their FUV$-V$ colors unusually blue for their metallicities.  
These are thought to be analogs of the two peculiar Galactic GCs 
NGC 6388 and NGC 6441 with extended blue HB stars. Based on the models 
incorporating helium enriched subpopulations in addition to the majority 
of the population that have a normal helium abundance, we suggest that 
even small fraction of super-helium-rich subpopulations in GCs can 
reproduce the observed UV bright metal-rich GCs. 
Young clusters in M31 show distinct UV and optical properties from 
GCs in Milky Way.  In general, young clusters are bluer than old GCs 
in both NUV and FUV.  Population models indicate that their typical 
age is less than $\sim$ 2 Gyrs and is consistent with the age derived 
from the most recent high-quality spectroscopic observations. 
A large fraction of young GCs have the kinematics of the thin, 
rapidly rotating disk component.   However, a subset of the old GCs also
shares the thin-disk kinematics of the younger clusters.
Most GCs with bulge kinematics show old ages.  The existence of young 
GCs on the outskirts of M31 disk suggests the occurrence of a 
significant recent star formation in the thin-disk of M31.  
Old thin-disk GCs may set constraints on the epoch of early formation 
of the M31 thin-disk. 
We detect 12 (10) intermediate-age GC candidates in NUV (FUV) 
identified by previous spectroscopic observations.  On the basis of 
comparing our UV photometry to population models, we suggest that 
some of spectroscopically identified intermediate-age GCs may not be truly 
intermediate in age, but rather older GCs that possess developed HB 
stars which contribute to enhanced UV flux as well as Balmer lines. 

\end{abstract}


\section{Introduction}

While the principle science goal of the {\sl Galaxy Evolution Explorer} 
({\sl GALEX}) has been the study of star formation in the local and intermediate 
redshift Universe, nearby galaxies such as M31 have also been surveyed, 
taking advantage of the wide (1.2 deg) field of view of {\sl GALEX}.   
Since launch in April 2003, we have been able to image 23 fields near and 
coincident with M31, now covering nearly all of M31 and its environs.  
Our first investigation of M31 globular clusters (GCs) using {\sl GALEX} 
was reported in Rey et al. (2005, hereafter Paper I). The dataset we report 
on here includes analysis of the coadded images, the wider field, and matching 
with optical/infrared datasets. While deeper observations of specific regions 
in M31 will likely occur in the next few years, it is unlikely that 
the combination of areal coverage and depth will be dramatically
superseded by any {\sl GALEX} observations in the next few years.  

GCs, as aggregates of coeval stars with small 
internal dispersion in chemical abundance, are the best examples of 
simple stellar populations and ideal templates for studying the 
evolution of stellar populations.  Observations of GCs offer the 
means of clarifying fundamental parameters in various astrophysical 
processes.   At present, GCs are being used to constrain the 
connection between GC systems and their host galaxies, since they 
provide tracers of the star formation histories of galaxies in the 
sense that major star formations in galaxies are concomitant with 
global GC formation (Brodie \& Strader 2006).  In this respect, 
relative ages of GCs within a galaxy or between different galaxies 
are crucial for understanding the time scale for galaxy formation 
and subsequent chemical evolution.  

The most direct way of measuring age differences among different 
stellar populations is to compare the observed luminosity levels of 
the main-sequence turnoffs (MSTOs); for a given metallicity an
intrinsically fainter MSTO translates to older ages.  Unfortunately
this method is limited to the nearest GCs, where individual
stars can be resolved and measured well to two magnitudes fainter than
the MSTO. Even in the nearest massive spiral galaxy, M31, the MSTO 
has been reached for only one cluster (G312=B379) and that only after 
$\sim 3.5$ days of $HST$ ACS integration time (Brown et al. 2004).

Alternatively, ages can be estimated from the horizontal-branch (HB) 
morphologies of GCs.  For a sufficiently old ($>$ 8 Gyr) stellar 
population, the surface temperature distribution of HB stars is 
sensitive to age (Lee, Demarque, \& Zinn 1994; Yi 2003).
Despite uncertainties of the details regarding peculiar HB 
morphologies in some Galactic GCs (e.g., bimodal HB distribution 
and extreme HB stars), various studies have suggested that together 
with metallicity, cluster age is the major parameter controlling 
the HB morphology (Lee et al. 1994; Sarajedini, Chaboyer, 
\& Demarque 1997; Rey et al. 2001; Salaris \& Weiss 2002).  

The hot He-burning HB stars and their progenies are most likely the 
dominant ultraviolet (UV) sources in the nearest old stellar populations 
(O'Connell 1999; Brown 2004). The integrated far-ultraviolet (FUV) flux depends 
mainly on the fractional number of HB stars with $T_{eff}$ hotter 
than $\sim 10,000$ K, with a modest dependence on their temperature 
distribution.  The production rate of these hot HB stars increases 
with age.  As a result, older GCs are likely to produce stronger 
FUV flux at a given metallicity (Park \& Lee 1997; Yi et al. 
1999; Lee, Lee, \& Gibson 2002). In this respect, integrated FUV flux 
can act as an age indicator of GCs (see also Dorman et al. 1995, 2003; 
Yi 2003; Catelan 2005; Kaviraj et al. 2006). Furthermore, this allows 
us to investigate mean age distributions of GCs within a single galaxy 
or among different galaxies, provided that other minor sources of FUV flux 
(e.g., existence of extreme HB stars due to the helium variation among 
stellar population in a GC; Lee et al. 2005 and references therein) 
can be quantified.

The M31 GC system (GCS) is one of the most ideal targets for 
integrated UV photometry due to its proximity (770 kpc; Freedman 
\& Madore 1990) and populous GCs (more than 300 confirmed GCs; Galleti 
et al. 2004).  Catalogs of M31 GCS are compiled from various 
imaging observations (Battistini et al. 1980, 1987, 1993; Crampton et al. 1985; 
Reed, Harris, \& Harris 1994; Barmby et al. 2000; Galleti et al. 2004).  
In addition, many optical spectroscopic observations 
provide chemical and kinematical properties of M31 GCS (Huchra, 
Stauffer, \& Van Speybroeck 1982; Kent, Huchra, \& Stauffer 1989; 
Huchra, Brodie, \& Kent 1991; Brodie \& Huchra 1991; Barmby et al. 
2000; Perrett et al. 2002).  There also have been color-magnitude 
diagram (CMD) studies of a few M31 GCs using data obtained by {\sl HST} 
(Ajhar et al. 1996; Fusi Pecci et al. 1996; Rich et al. 1996, 2005; 
Holland, Fahlman, \& Richer 1997; Jablonka et al. 2000; Stephens et al. 2001).  
All of these studies suggest that the M31 GCS is very similar to that 
of the Milky Way in many ways.  Therefore, the M31 GCs are
a kind of bridge between resolved and unresolved stellar populations, making
these clusters a critical set of template populations.

On the other hand, there has been growing body of evidence that 
the GCS of the Milky Way and of M31, the two large spirals in the 
Local Group, have very different properties and evolutionary 
histories (c.f. Rich 2004). In contrast to the case of the Miky Way 
halo GCs with uniformly very old (10 -- 12 Gyr) ages, there have been 
suggestions that the age distribution of M31 GCs is wide. There is a 
large population of young clusters with ages less than 1 -- 2 Gyr and 
intermediate-age GCs with $\sim$ 5 -- 8 Gyr (Burstein et al. 2004; 
Beasley et al. 2004, 2005; Fusi Pecci et al. 2005; 
Puzia, Perrett, \& Bridges 2005) are found in M31.  Furthermore, 
M31 possesses a kinematically distinct group of GCs that follows 
the disk rotation curve of M31 (Morrison et al. 2004), whereas no 
such counterparts have yet been discovered in Milky Way. 
Finally, the nitrogen abundance of the M31 GCs appears to be enhanced 
relative to that of the Milky Way GCs (Burstein et al. 2004). 

To what degree do the GCSs of M31 and the Milky Way share 
similarities?  This question can be answered by taking an approach 
different from previous studies, i.e. using UV photometry which is 
sensitive to age.  So far, there has been no systematic UV survey 
for a large and representative sample of GCs in galaxies, 
where detailed information on their optical and near-infrared 
photometry are available.  While there are some UV imaging 
studies of Galactic GCs from satellites such as the  
{\sl Orbiting Astronomical Observatory (OAO2)}, the 
{\sl Astronomical Netherlands Satellite (ANS)}, and the 
{\sl Ultraviolet Imaging Telescope (UIT)} (Dorman, O'Connell, 
\& Rood 1995 and references therein), very little have been 
carried out for M31 GCs.  Using the {\sl International 
Ultraviolet Explorer (IUE)}, Cacciari et al. (1982) and Cowley 
\& Burstein (1988) obtained the first spectra of a dozen 
brightest clusters in M31 at low resolution.  From rocket-borne 
UV imaging telescopes, near-ultraviolet (NUV) and FUV fluxes were 
determined for 17 cluster candidates in M31 (Bohlin et al. 1988).  
$UIT$ observations of M31 detected 43 GCs in NUV but only 4 
clusters in the FUV (Bohlin et al. 1993).  These studies 
show that the GCs of M31 and Milky Way cover a similar range of 
UV$-V$ colors.  Based on their UV results, Bohlin et al. (1993) 
suggest that the ages of M31 GCs are as old as the Galactic GCs.  
However, because the UV sample size of M31 GCs detected from 
$UIT$ is insufficient, especially in FUV which is more 
sensitive to the age of old stellar populations, the conclusion 
by Bohlin et al. can not be taken as definitive.  

Thanks to the successful launch and operation of {\sl GALEX}, 
the situation for UV studies of stellar clusters in nearby galaxies 
has dramatically changed. Using the unprecedented set of $GALEX$ 
photometry of M31 coupled with most recent detailed population models, 
we are now able to study the global UV characteristics of M31 GCs, 
and compare them to Milky Way GCs.  Using UV images obtained as part 
of the ``{\sl GALEX} Ultraviolet Survey of Globular Clusters in Nearby 
Galaxies" project, we present the catalog of {\sl GALEX} UV photometry of M31.  
This catalog is much more homogeneous than the Milky Way UV sample 
which was collected from several different UV satellites with 
heterogeneous selection criteria.  Unfortunately, {\sl GALEX} cannot study
a large fraction of Milky Way and Magellanic clusters because of
bright star avoidance restrictions. In this study, we focus on the 
interpretating the UV photometry of M31 GCs, combining existing 
optical photometry and corollary data from which we obtain insight 
into age distributions compared to the Galactic counterparts. 
In Paper I, we presented our first analysis of {\sl GALEX} photometry
of M31 GCs; this study concluded that the UV photometry of the older 
M31 GC is consistent with no age difference between M31 and the Milky Way.  
Our current paper supersedes that study both in the accuracy of the photometry
and in the numbers of detected M31 GCs.

In Section 2, we describe our observations and data analysis.  
Section 3 provides a comparison of age distribution of old GCs 
between M31 and the Milky Way.  In Section 4, we report the discovery 
of UV bright metal-rich GCs in M31, which are thought to be analogs 
of clusters with peculiar HB morphologies such as NGC 6388 and 
NGC 6441.  Section 5 describes the young clusters with $<$ 2 Gyr and 
the age distribution of thin-disk GCs found in M31.  In section 6, 
we discuss the intermediate-age GCs with 4 -- 7 Gyr identified by 
spectroscopic observations. The conclusions are summarized in Section 7.

\section{Observations and Data Analysis}
\subsection{Observations}

The M31 images were obtained as part of the Nearby Galaxy Survey  
(NGS) carried out by $GALEX$ in two UV bands: FUV (1350 -- 1750\AA)  
and NUV (1750 -- 2750\AA).  The data used in this study consist of  
images of 23 different 1.25 deg circular fields.  Details on the  
early observations of 14 fields released under the {\it GALEX Release 1 
(GR1)}\footnote{http://galex.stsci.edu/gr1} can be found in Thilker 
et al. (2005a) and Paper I.  In the 
interests of increasing both spatial coverage and depth around the 
M31 field, we have observed an additional nine $GALEX$ fields in 
September -- November, 2004, which are included in the {\it GALEX 
Release 2 (GR2)}\footnote{http://galex.stsci.edu/gr2}.  Our strategy 
for designing the pointings is to cover as wide area as possible 
around M31, but this was limited by the observing requirements of 
avoiding extremely UV-bright foreground stars.  The resulting 
coordinates and exposure times of the nine fields are listed in Table 1, 
and Figure 1 shows the field coverage of each 
$GALEX$ pointing around M31 projected onto 6\arcdeg$\times$6\arcdeg\ 
DSS images.  Our entire survey coverage of the 23 images is $\sim 17$ 
square deg.  Photon count maps from single visits were co-added to 
produce the final image for each field.  Preproccessing and calibrations 
were performed via the $GALEX$ pipeline.  Details of the $GALEX$ 
instrument and data characteristics are found in Martin et al. (2005) 
and Morrissey et al. (2005). 

\subsection{Photometry}

In a typical $GALEX$ image, point sources within 0.1 deg of the edge appear  
distorted due to the optics.  By examining each image, we decided to  
only use the inner 1.1 deg field, and pixels outside of this range  
were masked out prior to doing photometry.  Photometry of point 
sources in the M31 fields were carried out using the DAOPHOT II 
package (Stetson 1987).  For consistency, we have redone photometry 
for the 14 fields used in Paper I, although we later found that 
the new photometry is completely consistent with the old results 
within our errors.

The standard FIND-PHOTOMETRY-PICKPSF-PSF-ALLSTAR routine was applied  
to each image.  We then reran the PHOTOMETRY routine, this time  
taking outputs from the ALLSTAR program as inputs to ensure that the  
center of each source is better determined.  Visual inspection shows  
that with this method, centering is much better than solely running  
the PHOTOMETRY routine, especially for sources in crowded regions. 
The PHOTOMETRY routine was used to measure fluxes within a radius of 
3 pixel (4\farcs5) for each point source in both FUV and NUV images. 
Aperture corrections were derived using 19 -- 44 (5 -- 16) isolated  
stars per frame in NUV (FUV).  Mean aperture corrections for the  
entire data set are $-0.248 \pm 0.011$ and $-0.147 \pm 0.022$  
in NUV and FUV, respectively, where errors are standard deviations  
for the distribution of aperture correction.  Finally, flux  
calibrations were applied to bring all measurements into the  
AB magnitude system (Oke 1990; Morrissey et al. 2005). 
Astrometry for images processed through the $GALEX$ pipeline is known 
to be better than 1\farcs2 for 80\% of the stars in the entire frame
(Morrissey, private communication).
Our test shows that all stars common in both our photometric  
catalog and the USNO-B1.0 catalog (Monet et al. 2003) match to  
within 1\farcs5. 

\subsection{The Ultraviolet Catalog of M31 Globular Clusters}

The analysis presented in this work is based on an optically selected  
sample of GCs and GC candidates extracted from existing catalogs. 
The most comprehensive catalog of M31 clusters currently available 
is the Revised Bologna Catalog (RBC; Galleti et al. 2004).  This 
catalog provides positional and photometric information.  Out of 
the total entries in the RBC, we consider 1035 objects consisting of 
337 confirmed GCs, 688 GC candidates, and 10 objects with 
controversial classification.  The total number of RBC objects 
falling in the area covered by our $GALEX$ observations is 989. 

Sources in our $GALEX$ photometry were cross-matched with the RBC 
objects in the following manner.  We decided to use a matching  
radius of 6\arcsec considering the combined effect of the astrometry 
error in RBC ($< 1$\arcsec; see Fig 1 of Galleti et al. 2004) and 
$GALEX$ images ($< 1$\arcsec.2), and the 4\farcs6 FWHM PSF.  For 
each $GALEX$ image, we first generate a list of UV-detected RBC 
sources.  We then visually inspect each matched RBC source in 
each $GALEX$ image and reject all spurious sources.  Out of the 
total detected RBC sources, 38\%\ and 44\%\ in each NUV and FUV 
band were rejected by visual inspections.  Finally, we construct a $GALEX$ 
catalog with all UV detected RBC objects. 

NUV and FUV magnitudes and their associated errors for each object 
are listed in Columns (7)--(10) of Table 2.  As shown in Figure 1, 
each $GALEX$ image overlaps with at least another image.  For 
objects that lie in these overlapping fields, we have listed their 
mean magnitudes weighted by their photometric errors.  In total, 
our $GALEX$ UV catalog includes 485 and 273 RBC objects in NUV and 
FUV, respectively.  Columns (1) and (2) of Table 2 are identification 
of M31 GCs adopted from RBC and Barmby et al. (2000), respectively. 
In column (3), we also add the identification listed in NASA/IPAC 
Extragalactic Database (NED).
Column (4) is the classification flag describing the nature of the 
entry as listed in RBC; 1: confirmed GCs, 2: candidate clusters,
3: uncertain candidates.  RA and DEC coordinates, and $UBVRIJHK$ 
photometry in the table, i.e. columns (5)--(6) and (11)--(18), are 
also from the RBC.  Metallicities ([Fe/H]) and their errors in 
columns (19) and (20) are adopted from Barmby et al. (2000), while 
kinematic residuals in column (21) are taken from Morrison et al. 
(2004).  Figure 2 shows the spatial distribution on the 
sky of all RBC objects detected in $GALEX$ NUV and FUV bands with 
respect to the M31 disk, NGC 205, and M32. 

Extinction correction (including foreground and internal to M31) 
for each cluster is applied based on a list of E($B-V$) kindly  
provided by P. Barmby.  We use the reddening law of Cardelli, 
Clayton, \& Mathis (1989) to estimate the following; $R_{NUV}$ = 8.90 
and $R_{FUV}$ = 8.16.  In subsequent analyses, we only use confirmed 
M31 GCs [class 1 in column (4) of  Table 2] with E($B-V$)$<$0.16 
unless otherwise noted.  This limit corresponds to the median value 
of the E($B-V$) distribution for M31 GCs (Barmby et al. 2000).   

\subsection{$GALEX$ UV Detection Rate}

We use the $B$ and $V$ magnitudes of RBC to examine the detection 
rate of clusters in our $GALEX$ fields. Figure 3 shows the fraction 
of RBC objects detected in the NUV and FUV bandpasses as a function
of $V$ magnitude and $B-V$ color.  Of the 693 objects with both $B$ 
and $V$ data in RBC, 485 (about 70\%) and 273 (about 39\%) objects 
have their entries in the $GALEX$ UV catalog for NUV and FUV, respectively 
(Table 2).  The CMD and color histogram show that most of the detected 
objects belong to an optically blue subgroup with $B-V < 1.2$. It is 
interesting to note that many of the bluest clusters with $B-V < 0.5$ 
are detected in the $GALEX$ UV bands even though they are fainter than the 
redder clusters in the optical passband.  Most of these blue clusters 
are young clusters (see Section 5). We estimate that the limiting 
magnitudes are $\sim 22.5$ in both NUV and FUV. 

\subsection{UV$-$Optical Color Distribution}

Figure 4 shows the extinction-corrected $(NUV-V, M_{V})$ and  
$(FUV-V, M_{V})$ CMDs for M31 clusters ({\it large circles}) and 
Milky Way GCs ({\it crosses}).  The colors and magnitudes of Milky 
Way GCs were taken from Table 6 of Sohn et al. (2006), and the 
corrections for difference in effective wavelength of separate 
observations were made using the same method described in Paper I.  
A distance modulus of $(m - M)_0 = 24.43$ (Freedman \& Madore 1990) 
was adopted for all M31 clusters.  We also plot clusters ({\it small 
circles}) with no individual reddening information available in the 
literature, assuming that they are only affected by the foreground 
Galactic reddening of E($B-V$) = 0.10 (Crampton et al. 1985). 

The UV-detected M31 clusters essentially cover the entire $V$-band 
luminosity range present in the UV-detected sample of Milky Way GCs.  
The most distinct feature of Fig. 4 is that M31 clusters can be 
divided into two subgroups based on their UV$-V$ colors; UV-red GCs 
with $(NUV-V)_0 \ga 3.0$ and $(FUV-V)_0 \ga  4.0$, and UV-blue GCs 
with $(NUV-V)_0 \la 3.0$ and $(FUV-V)_0 \la 4.0$.  The red subgroup 
of M31 clusters and Milky Way GCs generally occupy the same area in 
the CMDs.
Also, most M31 clusters that belong to the red subgroup have UV$-V$ 
colors [$(NUV-V)_0 < 4.3$; $(FUV-V)_0 <5.5$] consistent with 
theoretically-predicted colors of the UV bright clusters with blue 
HB morphology (see Section 4).  M31 clusters with $(NUV-V)_0 > 4.3$ 
are likely to have redder HB morphologies and relatively faint UV 
luminosities. 

The blue subgroup of M31 clusters is significantly bluer than the 
Milky Way GCs.  Among their $UIT$ sample, Bohlin et al. (1993) also  
reported clusters much bluer than Milky Way GCs in the NUV.   
They suggest that these clusters are significantly younger than  
typical GCs of $> 10 Gyr$ (see also Burstein et al. 1984; Bohlin et 
al. 1988; Cowley \& Burstein 1988).  Many blue clusters included 
in our catalog have colors consistent with those [$(NUV-V)_0 \la 2.5$] 
of younger clusters suggested by Bohlin et al. (1993), while no Galactic 
counterparts are found.  Most of them show stronger fractional UV 
flux than those of old red GCs, which is suggestive of young ages.
We discuss the details of these clusters in Section 5.

\section{Age Distribution of Old Globular Clusters}

Various studies suggest that UV$-$optical colors can be an age-dating 
tool for old stellar systems (Lee et al. 2002; Yi 2003; Paper I; 
Catelan 2005). Specifically, Lee et al. (2002) and Yi (2003) show that 
FUV$-V$ color can be more sensitive to ages than any optical-band 
integrated color.  While the strength of FUV flux depends 
on the fractional number of hot HB stars, such stars 
contribute to little light to the optical bands, that are instead 
dominated by MS and giant stars (Dorman et al. 1995).  At a given 
metallicity, older populations with bluer HB stars should be 
relatively UV bright, especially in the FUV passband.  Thus, the 
FUV$-V$ color can be a measure of the age variation based on the 
fraction of the stellar population that appears to be hot HB stars. 

Figure 5 shows the [Fe/H] vs. UV$-V$ diagrams for M31 clusters 
({\it circles}) and Galactic GCs ({\it crosses}), as an analog of the [Fe/H] vs.
HB type diagram of GCs (Lee et al. 1994; as such, bluer colors are 
now on the right). We restrict the M31 sample to the supposedly old, 
red clusters with E($B-V$) $<$ 0.16. The [Fe/H] values for M31 is 
taken from Table 2, while for the Galactic GCs, we adopt 
the February 2003 version of Harris (1996) catalog
\footnote{http://physwww.mcmaster.ca/\%7Eharris/mwgc.dat}.
While the M31 clusters extend over a wide range of $(NUV-V)_0$ color
comparable to that of Galactic GCs, 
the FUV sample is biased to the blue and metal-poor GCs with no redder 
[$(FUV-V)_0 > 6.5$] clusters due to their instrinsically faint FUV 
brightnesses.

The model isochrones in Fig. 5 are constructed from our evolutionary 
population models of GCs in the $GALEX$ filter system 
(Chung, Yoon, \& Lee 2006, in preparation; see also Lee et 
al. 2002; Yi 2003). The models include the treatment of the detailed 
systematic variation of HB morphology with age and metallicity and 
the fractional contribution of post-asymptotic giant-branch stars.
The $\Delta t$ = 0 ({\it solid line}) isochrone corresponds to inner halo 
Galactic GCs (Galactocentric radius $\leq$ 8 kpc) of $\sim 12$ Gyr.  
The $\Delta t$ = +2 Gyr ({\it long dashed line}) and 
$-2$ Gyr ({\it dotted line}) isochrones are for the models 2 Gyr 
older and younger than the inner halo Galactic GCs, respectively. 

For a given age, HB morphology gets bluer with decreasing metallicity 
and the population models produce increasing UV flux.  
As shown in the left panel of Fig. 5, M31 and Galactic GCs follow 
the general trend indicated by the isochrones. At a fixed metal abundance, 
the models exhibit rapid blueing of FUV$-V$ colors as the age increases.  
The large separation among isochrones with different ages 
(see right panel of Fig. 5) confirms that FUV$-V$ color is more sensitive 
to ages than any optical-band integrated color.  The NUV$-V$ is relatively 
insensitive to age compared to FUV$-V$ for old ($\ga$ 8 Gyr) GCs. This is because
not only HB stars but also MSTO stars are responsible for producing 
integrated NUV flux (Lee et al. 2002; Dorman et al. 2003).  
The model isochrones fit well the range in color of the GCs and their  
general locus on the age-metallicity plot.  In the right panel of Fig. 5, 
M31 and Galactic GCs are well located in the age range ($\pm2$ Gyr) 
of model predictions and show no significant difference in age distribution 
between the two GCSs.  This indicates that M31 and Milky Way GCSs 
share similarities in mean age and age spread, at least, for the 
old ($\ga$ 8 Gyr) GCs.

Many M31 GCs are resolved by $HST$ and for those we get HB 
morphology directly, which makes possible a comparison of the 
integrated properties of GCs with their actual resolved stellar 
populations (e.g., Rich et al. 2005).  Based on the previous $GALEX$ 
UV data for M31 GCs with known HB morphology from $HST$ observations, 
Paper I confirms that the sensitivity of metallicity and HB 
morphology to the UV flux is evident in both NUV and FUV bands. 
Recently, Brown et al. (2004) performed a direct age estimation 
of the M31 GC B379 by fitting its deep $HST$ ACS photometry extending 
below the MSTO.   The extreme stellar crowding caused by the compactness
of the cluster results in a less well determined CMD, leading to a 
somewhat weak age constraint 10$^{+2.5}_{-1}$ Gyr. Nonetheless, 
they suggest that B379 appears to be 2 -- 3 Gyr younger than Galactic 
GCs with similar metallicities. The CMD of B379 shows a tight clump 
of red HB population consistent with its metallicity and suggested age 
(Brown et al. 2004; Rich et al. 2005).  We detect this cluster 
({\it filled circle}) only in the NUV passband and its $(NUV-V)_0$ 
color is in good agreement with its HB morphology.

Due to the increased depth of our UV images compared to Paper I, the 
number of NUV- and FUV-detected GCs with [Fe/H] $> -1$ has grown. 
It is interesting to note that metal-rich GCs have a wide dispersion 
of UV colors [$(NUV-V)_0$$\sim$ 4 -- 6] in the [Fe/H] vs. $NUV-V$ 
plot.  Furthermore, although the  $NUV-V$ color is relatively less 
sensitive to the age variation, metal-rich GCs appear to distribute 
tightly around isochrone of $\Delta t$ = +2 Gyr.  Since there is no 
clear evidence that the metal-rich GCs in M31 are systematically 
older than metal-poor counterparts (e.g., Jiang et al. 2003; Puzia 
et al. 2005), this feature may reflect the variation of HB morphology 
and UV flux of M31 metal-rich GCs due to parameters other than 
metallicity and age (see Sec. 4 for details).  Contrary to the case 
for most Galactic GCs with [Fe/H] $\sim$ $-0.4$ -- $-0.8$, many 
metal-rich M31 GCs show bluer $NUV-V$ colors (i.e., stronger 
UV flux).  Even some metal-rich GCs show comparable $NUV-V$ 
colors to those of intermediate metallicity GCs with [Fe/H] 
$\sim$ $-1$ -- $-1.5$. Only two peculiar metal-rich Galactic GCs, NGC 
6388 and NGC 6441 have NUV colors comparable to those of metal-rich 
M31 GCs.  Among the metal-rich M31 GCs, three GCs are also detected in FUV. 
Most metal-rich GCs detected in NUV are considered to have blue HB 
stars which contribute flux in NUV passbands, whereas three 
metal-rich GCs detected in FUV would have more extreme hot blue HB 
populations (e.g., NGC 6388 and NGC 6441 in the Milky Way; see Sec. 4 
for details).  In this respect, bright UV flux in M31 metal-rich GCs 
is in good agreement with the explanation that stronger H${\beta}$ 
lines observed in M31 metal-rich GCs result from the presence of 
old blue HB stars (Peterson et al. 2003) instead of them being 
significantly younger than the bulk of M31 GCs (Burstein et al. 1984).

\section{UV Bright Metal-Rich Globular Clusters}

Galactic GCs such as NGC 1851 and NGC 2808 exhibit bimodal HB 
distributions, where separate group of hot HB populations coexist 
with cooler HB components.  The most striking cases of GCs with HB 
bimodalities are NGC 6388 and NGC 6441.  These two have pronounced 
hot HB populations that are not seen in any other Galactic GC of 
similar metallicity.  GCs with [Fe/H] $> -0.8$ (e.g. 47 Tuc) are 
typically faint in the UV due to their HB being dominated by cool 
stars.  Rich, Minniti, \& Liebert (1993), however, detected 
abnormally strong UV flux in NGC 6388 and NGC 6441 from IUE 
observations.  Rich et al. (1993) suggest the existence of hot 
extreme HB stars in these clusters.  The $HST$ observations have 
directly reveal a prominent extended blue HB population, in 
addition to a well populated red HB sequence (Rich et al. 1997; 
Piotto et al. 1997).  While the cause of peculiar blue HBs in NGC 
6388 and NGC 6441 is still not settled, the existence of hot HB 
stars in metal-rich GCs has implications for understanding the 
evolution of GCs and, furthermore, elliptical galaxies, since it is 
related to the origin of UV upturn phenomenon (O'Connell 1999 and 
references therein).

While NGC 6388 and NGC 6441 stand out as the only two metal-rich 
GCs with prominent blue HB populations in the Milky Way, it is 
intriguing to search for examples of extragalactic counterparts to see
whether these objects are ubiquitous.  UV observations provide an
indirect solution for this purpose since integrated UV flux is 
significantly sensitive to the presence of hot stars in evolved 
stellar populations.  In our previous study (Paper I), we have 
detected seven clusters with [Fe/H] $> -1$ in the NUV but not 
in the FUV.  With our deeper $GALEX$ observations, we are now better able 
to search M31 for counterparts of NGC 6388 and 6441.

Figure 6 is the same as Fig. 5 but now we highlight UV bright 
metal-rich GCs in M31.  In our $GALEX$ M31 sample, we find three 
candidate UV bright metal-rich GCs ({\it filled circles}: B131, 
B193, and B225) with E($B-V$)$<$0.16.  These have $FUV-V$ 
colors similar to or smaller than those of NGC 6388 and NGC 6441. 
Fortunately, CMDs exist for cluster B225 based on images obtained 
with $HST$ WFPC2 (Fusi Pecci et al. 1996; Rich et al. 2005) and 
NICMOS (Stephens et al. 2001).  The $BV$ CMD of B225 (see Fig. 4 
of Fusi Pecci et al. 1996; see also Fig. 9 of Rich et al. 2005) show 
a hint of blue HB population around $(B-V)_{o}$ = 0 in addition to 
the red HB population.  Overall, the HB morphology appears to be 
similar to those of NGC 6388 and NGC 6441, although the existence of 
an extended blue HB sequence in B225 is difficult to confirm because
the CMD reach only 1 mag fainter than the HB level.  The [Fe/H] 
values determined from the photometry by Fusi Pecci et al. (1996) and 
Rich et al. (2005) are also similar to those for NGC 6388 and NGC 6441. 
Deep imaging with $HST$ is required to judge whether the UV bright 
metal-rich clusters (including B225) are M31 counterparts of 
NGC 6388 and 6441.

Using the Ca II index as an indicator of HB morphology (Rose 1984, 
1985), Beasley et al. (2005) suggest that two M31 GCs (B158 and 
B234) are candidate metal-rich GCs that may have blue HB 
components and therefore are analogs of NGC 6388 and NGC 6441.  
The Ca II indices of these two M31 GCs are similar to those of NGC 
6388 and NGC 6441 and are systematically smaller than those of other M31 
and Galactic GCs with similar metallicity (see Fig. 19 of Beasley 
et al. 2005). We detected B158 ([Fe/H] = $-1.08$) in both NUV and 
FUV, but B234 ([Fe/H] = $-0.84$) is detected only in NUV ({\it filled 
triangles} in Fig. 6).  As shown in Fig. 6, cluster B234 shows 
rather larger NUV flux than those of NGC 6388 and NGC 6441.  On the 
other hand, cluster B158 shows normal UV colors in both NUV 
and FUV which are consistent with its relatively low metallicity.

In Fig. 6, we include the extremely massive and moderately 
metal-rich ([Fe/H] = $-1.08$; Barmby et al. 2000) M31 GC G1 
({\it filled square}).  The CMD of G1, as presented by Rich et al. (2005),  
is similar to that of 47 Tuc; the HB is dominated by cool stars.  
%This suggests that G1 and 47 Tuc have similar ages and metallicities. 
However, G1 shows a notable minority population of blue HB stars near 
the detection limit, which is consistent with slightly stronger 
UV flux than GCs with similar metallicity (see also Peterson et al. 2003). 
Therefore, G1 is considered to be a moderate metal-rich GC with 
low-level enhancement of UV flux.

According to its HB morphology in the {\it HST} CMD, cluster B225 is 
almost certainly an old GC.  However, since young populations can also 
contribute to the integrated UV flux (see also Sec. 5), we need to check 
if the other two clusters B131 and B193 are similarly old. In Fig. 6, 
we superpose a 1 Gyr isochrone ({\it short dashed line}) in addition 
to three isochrones of age 10 -- 14 Gyr.  The UV bright metal-rich clusters 
in M31 and NGC 6388/NGC 6441 sit well between the 1 and 14 Gyr isochrones 
making the age discrimination ambiguous.  However, using the color-color 
diagram such as $B-V$ vs. $FUV-V$ as shown in Figure 7, it is possible 
to separate young clusters from old GCs.  For example, young clusters are 
bluer than old GCs in both $B-V$ and $FUV-V$ for the metallicity range of 
[Fe/H] $<$ -1.0. Metal-rich ([Fe/H] $>$ -1.0) young GCs in the range of 
$(FUV-V)_{o}>$ 4.0 are also reasonably deviated from the old GCs. 
All candidate UV bright metal-rich GCs in M31 follow the isochrones for 
old ages, which confirm that they are bona-fide old metal-rich GCs with 
strong FUV flux. 

Recently, Sohn et al. (2006) obtained {\sl HST} STIS FUV and NUV 
photometry of 66 GCs in the giant elliptical galaxy M87.  
A remarkable and unanticipated result is that M87 GCs have distinct 
UV properties from GCs in the Milky Way and M31, despite strong 
overlap in optical properties.  Their $FUV-V$ colors are up to 1 mag 
bluer than any Milky Way GC. M87 GCs appear to produce larger hot HB 
populations and therefore much more UV flux than do Milky Way and 
M31 counterparts at a given metallicity.  However, there is no ready 
explanation for the prominent strength of hot HB stars in the M87 GCs 
(see Sohn et al. 2006 for details).  In [Fe/H] vs. UV$-V$ diagrams of 
Figure 8, we compare GCs in M87 ({\it filled squares}; Sohn et al. 2006) 
with those in Milky Way ({\it crossess}) and M31 ({\it open circles}).  
We convert magnitudes of M87 GC data in STMAG system to those of the 
$GALEX$ filter system.  The [Fe/H] values of M87 GCs are transformed 
from $V-I$ (Sohn et al. 2006).  Most M87 GCs are systematically 
brighter in UV than the Milky Way and M31 GCs at any given [Fe/H].  
However, it is worth noting that the distributions of UV bright 
metal-rich GCs in M31 ({\it filled circles}) and NGC 6388 and NGC 6441 
are in good agreement with that of metal-rich M87 GCs in [Fe/H] vs. 
($FUV-V$)$_0$ diagram.  This indicates that the UV-detected GCs in 
M87 may be mostly analogs of NGC 6388 and NGC 6441.  The UV bright 
metal-rich GCs in M31 and Milky Way constitute a distinct type of 
stellar population and share their properties with metal-rich GCs 
in M87.  The inference is that these anomalous UV bright metal-rich 
GCs may be ubiquitous in more or most galaxies, although M87 is 
distinct from the other two.

While we still lack a definitive explanation, several scenarios have 
been suggested to elucidate the observed peculiar HB morphology and 
RR Lyrae periods in the NGC 6388 and NGC 6441 (see Catelan 2005 
for details and references therein).  A recently revived possibility 
is that helium abundance may play an important role for the existence 
of the hot extreme HB stars and for pronounced UV flux of metal-rich 
GCs.  S.-J. Yoon et al. (2006, in preparation) discuss the possibility  
that the bimodal nature of the HBs of NGC 6388 and NGC 6441 can be 
attributed to a small fraction of super-helium-rich (Y $\sim$ 0.3) 
population due to the self-helium-enrichment (see also Catelan 2005;
Moehler \& Sweigart 2006). 
Considering the CMD morphologies and the RR Lyrae properties, they 
conclude that the stellar population simulations based on the 
super-helium rich hypothesis reproduce the observations more 
successfully for these GCs.  Meanwhile, it is also proposed that the 
subpopulations of two other Galactic GCs, $\omega$ Centauri and NGC 
2808, are super-helium-rich (D'Antona \& Caloi 2004; Lee et al. 
2005; D'Antona et al. 2005; Bekki \& Norris 2006).  In the case of $\omega$ Cen,
the helium rich population actually occupies a distinct main sequence locus.
Furthermore, combining $HST$ STIS (Sohn et al. 2006) and 
WFPC2 (Jordan et al. 2002) photometry with stellar population models, 
S. Kaviraj et al. (2006, in preparation) propose that the majority of 
UV-bright M87 GCs show strong signatures of minor super-helium-rich 
populations.  All of these results suggests that the HB morphology 
is also governed by the helium abundance variation among the stellar 
populations in some GCs as a possible third parameter.

Figure 9 is a comparison of UV bright metal-rich GCs in Milky Way 
(NGC 6388 and NGC 6441), M31 ({\it filled circles}), and M87 
({\it filled squares}) with our model isochrones overlaid. 
Dotted lines are for the case in which all the GCs have same primordial 
helium abundance of Y = 0.23 but different ages.
Dashed lines are for the case with minority population of helium 
enhancement (Y = 0.33) in addition to the majority of the population 
with normal helium abundance (Y = 0.23) within a model GC at the 
same age of 12 Gyr. These correspond to models that include the 
extreme HB stars with T$_{eff}$$>$15,000 K due to the helium enhancement
(see Lee et al. 2005; D'Antona et al. 2005). 
Different model lines with high helium abundances indicate different 
population number fraction of helium enhancement (left to right, 
10\%, 20\%, 30\%, and 50\%) within a GC.  Strong UV flux can be 
reproduced by even small fraction of helium enhanced subpopulation 
within a GC.  A reasonable agreement between observed UV colors 
of UV bright metal-rich GCs and models is obtained when we 
adopt a wide range of helium abundance. One of the possibilities for 
the strong UV flux 
in metal-rich GCs may be understood as being due to the contribution of extreme 
hot HB stars in these clusters as a consequence of the large 
differences in helium abundances that easily overcomes the 
metallicity effect.

\section{Young and Thin-Disk Globular Clusters}
\subsection{Young Globular Clusters}

Young star clusters with mass range 10$^{4}$ -- 10$^{6}$ M$_{\odot}$ 
has been identified in a few distant galaxies including Local Group 
members (Larsen \& Richtler 1999).  Observations for such objects 
may provide valuable information concerning the formation and 
evolution of old GCs in galaxies. M31 is not exception in having 
these kinds of clusters that have similar luminosity 
and structure as the Milky Way old GCs, but show significantly bluer 
[$(B-V)_0 <$ 0.45; Fusi Pecci et al. 2005] integrated optical colors. 
Interestingly, in the Local Group, the Milky Way appears to be the 
large galaxy that lacks such young massive compact clusters. 
There are some fragmentary studies of young clusters in M31 
(Fusi Pecci et al. 2005 and references therein).  Early studies 
report the existence of these blue clusters in M31 (Vetesnik 1962; 
van den Bergh 1967, 1969; Searle 1978), and recent studies have 
concentrated on the properties of these objects 
(Williams \& Hodge 2001a,b; Beasley et al. 2004; 
Burstein et al. 2004; Puzia et al. 2005; Fusi Pecci et al. 2005).    

In the context of understanding the spectral evolution of simple 
stellar populations of star cluster, the integrated spectrum of young 
cluster is dominated by short-lived, young massive stars with strong 
UV flux (see Fig. 9 of Bruzual and Charlot 2003).  Therefore, 
compared to the optical passband, the UV is a good probe to identify 
young clusters.  In previous UV observations using rocket-borne 
telescope and $UIT$, only a handful of young GC candidates in M31 were 
identified (Bohlin et al. 1988, 1993).  Based on their blue UV 
colors, Bohlin et al. suggested that these are clusters with ages 
less than 1 Gyr that may be analogous to the large populations of 
``blue globulars" in LMC.

The combination of the UV$-V$ color with the $B-V$ color clearly 
improves the discrimination between old and young GCs.  This is  
based on a comparison of the slope of the spectral energy 
distribution in the $B-V$ range with the slope from the UV to $V$ 
band with distinctly different ages (Yi 2003).  Figure 10 shows 
$B-V$ vs. UV$-V$ diagrams for young GCs (B210, B222, B315, B321, 
B322, B327, B484, and V031; {\it filled circles}) found from the 
recent high-quality spectroscopic observations (Beasley et al. 2004; 
Burstein et al. 2004), compared with the old GCs ({\it open circles}).  
Among the young GCs in the list of Beasley et al. (2004), we exclude 
the slightly older ($\sim$ 2 -- 3 Gyr) metal-poor cluster B292.  
The mean distribution of young GCs is biased to bluer colors in both 
optical and UV colors.  Especially, in the right panel of Figure 10, 
young GCs are clearly separated from the old GCs.  Since the reddening 
values of these young GCs are not available from Barmby et al. (2000), 
we adopt the mean foreground reddening value of M31 field, E($B-V$) = 0.10.  
Ages of these young GCs are estimated to be in the range 0.5 Gyr 
({\it long dashed line}) -- 1 Gyr ({\it solid line}) according to our 
population models, which are consistent with results of Beasley et al. (2004) 
and Williams \& Hodge (2001a). 

One caveat for the studies of young GCs is that some of the 
suggested very young GCs in distant spiral galaxies may be 
spurious identifications rather than bona-fide clusters. 
Most recently, from their high spatial resolution observations 
with Keck laser-guide star adaptive optics system, Cohen et al. 
(2005) found that a large fraction of the putative very young 
($<$ 2 Gyr) GCs in M31 may not be genuine GCs, but probably 
asterisms.  If this is true, unless we undertake a similar survey 
for asterisms and identify true young GCs, we could not provide the 
exact frequency of young GCs, which is important to constrain the 
formation rate of GCs.  Four objects (B223, B216, B314, and B380; 
{\it filled triangles}) identified as asterisms by Cohen et al. (2005) 
are found in our UV catalog. They are also included in the list of 
young GCs of Beasley et al. (2004) and Burstein et al. (2004).  
As shown in Fig. 10, these objects have similar optical/UV colors 
to those of other possible young GCs; the UV photometry cannot
be used to distinguish true clusters from asterisms.

Meanwhile, Williams \& Hodge (2001a) presented $HST$ images for 
four young genuine M31 GCs that have the appearances of typical 
GCs (see Fig. 1 of Fusi Pecci et al. 2005).  CMDs with resolved 
stellar populations of these clusters yield ages less than 200 Myr. 
Four young GCs (B315, B319, B342, and B368; {\it open squares} 
in Figure 10) of Williams \& Hodge (2001a) are identified 
in both NUV and FUV passbands. Although Williams \& Hodge (2001a) 
estimated E($B-V$) values of these clusters using the 
color of the upper MS stars of CMDs, we adopt E($B-V$)= 0.10 as 
their reddening values for consistency. The $B$ and $V$ magnitudes 
are from the Table 1 of Williams \& Hodge (2001a).  Their UV 
colors are fully consistent with those of model predictions with 
young ages.  The integrated luminosities and half-light radii of 
these clusters indicate that they are more massive and compact 
than counterparts in Milky Way.  Furthermore, their luminosities 
are quite similar to young clusters found in many other spiral 
galaxies (Larsen \& Richtler 1999; Larsen 2004).  Evidently, 
these massive and compact young clusters are common in galaxies 
and/or, at least, in the Magellanic Clouds and M31; in more distant
galaxies, the problem of confusion with individual stars and 
clusters will be a more significant issue, however.

\subsection{Thin-Disk Globular Clusters}

In addition to the detection of many young GCs in M31 from various 
observations, Morrison et al. (2004) argues for the presence of a 
subsystem of GCs in M31 with thin-disk kinematics, based on the 
spectroscopic data of Perrett et al. (2002).  Thin-disk GCs are 
identified as clusters with residual radial velocity ($\delta$) 
less than 0.75 km s$^{-1}$ (see Morrison et al. 2004 for details).  
This new population of GCs is found across the entire disk of 
M31 and has a wide range of metallicities ranging from [Fe/H] $< -2.0$ 
to above solar, indistinguishable from M31 GC population. 
Lacking a direct measurement of the ages of these GCs, Morrison 
et al. (2004) speculate that M31 had a relatively large thin-disk 
during its early evolution, because these clusters appear 
to be old and metal-poor.  Furthermore, they suggest that the 
existence of such a disk with cold kinematics is not attributable to a 
substantial merger event or significant perturbation since the 
clusters were formed.  This is in contrast to the suggestion that 
M31 was formed by an equal-mass merger 6 -- 8 Gyr ago (e.g., 
Brown et al. 2003).

Subsequent studies of the disk clusters (Burstein et al. 2004; 
Beasley et al. 2004; Puzia et al. 2005) have reported that some 
thin-disk GC samples of Morrison et al. (2004) are massive young GCs. 
The recent extensive study by Fusi Pecci et al. (2005) found a 
population of 67 massive blue clusters in the disk of M31.  These 
clusters have very blue colors and enhanced H${\beta}$ absorption 
lines indicating an age of $\sim$ 2 Gyr, and possess kinematics 
similar to that of the M31 thin-disk.  Fusi Pecci et al. (2005) 
confirmed that these massive blue thin-disk clusters appear to 
avoid the inner regions of the M31, but are rather well projected 
onto the outer disk and coincide with continuous star-forming 
regions in the spiral arms.  These results lead to the suspicion
that a significant population of old metal-poor clusters 
exists in the thin-disk of M31 as suggested by Morrison et al. (2004).

In order to constrain the formation epoch and evolution of the M31 
thin-disk and the clusters in it, further studies of the age distribution 
with larger sample of thin-disk GCs are required.  Specifically, it 
is crucial to explore among clusters with thin-disk kinematics, (1) 
the number of young clusters and (2) whether there exists any old GC.  
At the same time, it is also useful to compare the age distribution 
of thin-disk GCs with that of GCs not showing thin-disk kinematics 
(i.e, bulge GCs).  For these purposes, we use $GALEX$ UV data of M31 
GCs to investigate the age distribution of subgroups of GCs with 
different kinematic properties. 
  
Figure 11 shows the $B-V$ vs. UV$-V$ color-color diagrams for M31 
GCs with thin-disk kinematics ({\it filled circles}).  As suggested by 
Morrison et al. (2004), GCs with absolute residual radial velocity 
less than 0.75 km s$^{-1}$ are selected as thin-disk GCs.  The dashed 
horizontal line corresponds to a reference color value [i.e., 
($B-V$)$_{o}$ = 0.45] for selecting young clusters following Fusi 
Pecci et al. (2005).  Their locations are consistent with those of 
confirmed young clusters (Figure 10) and our population models with 
less than 1 Gyr. It is evident that most candidate young GCs with 
blue optical colors [$(B-V)_0 <$ 0.45] show thin-disk kinematics. 
At least, about 70\% of young GC sample with $(B-V)_0 <$ 0.45 
detected in FUV falls into the thin-disk subsystem.

Besides GCs with known kinematic information from Morrison et al. 
(2004), some blue GCs without kinematic data [open circles with 
$(B-V)_0 <$ 0.45] also show young ages.  According to the age 
distribution of GCs with bulge kinematics (see Figure 12), 
they may be also possible thin-disk young GCs, but radial velocity
measurements for these clusters are needed.

It is worth noting that among old red GCs with $(B-V)_0 >$ 0.45, 
some thin-disk GCs are also found, although their fraction is not 
large.  This indicates that thin-disk GCs display large age spread, 
which is in good agreement with the results of Puzia et al. (2005). 
In turn, this suggests that the M31 thin-disk and clusters following 
its kinematics are formed at the early formation stage of M31, and 
that later on, a significant number of clusters are formed in the 
disk of M31. 

For comparison, in Figure 12, we plot M31 GCs with bulge kinematics 
({\it filled circles}) which are selected from the Morrison's $\delta$ 
larger than 0.75 km s$^{-1}$. Contrary to the thin-disk GCs, 
the most striking feature is that they are biased to the relatively 
red optical and UV colors.  This indicates that GCs with bulge kinematics, 
in the mean,  belong to the old subsystem. 

In Figure 13, we compare the distribution of M31 thin-disk GCs 
({\it filled circles}) with that of bulge GCs ({\it filled squares}). 
We divide thin-disk GCs into two groups:  blue [$(B-V)_0 <$ 0.45; 
{\it large filled circles}] and red [$(B-V)_0 >$ 0.45; {\it small 
filled circles}] GCs. Since [Fe/H] values of Barmby et al. are not available 
for many thin-disk GCs of Morrison et al. (2004), we instead adopt [Fe/H] 
from Perrett et al (2002).  In addition to the model isochrones with 
old ages comparable to mean Galactic halo (10 -- 14 Gyr, {\it dotted 
lines}), we also superpose our young isochrones with 1 Gyr ({\it solid 
line}) and 0.5 Gyr ({\it long dashed line}).  Optically blue thin-disk 
GCs show systematically stronger UV flux than do red thin-disk GCs at 
a given metallicity.  This again confirms that the majority of blue 
thin-disk GCs are young with ages less than $\sim 1$ 
Gyr. GCs with bulge kinematics are biased to the red UV color 
distribution, which is an indication of their old ages.

Besides the age distribution, the metallicity of young thin-disk GCs is 
another important clue to understand the formation and evolution of 
the thin-disk of M31.  In Fig. 13, the UV bright and young thin-disk 
GCs appear to be systematically more metal-poor than their old counterparts.  
Morrison et al. (2004) speculate that the large disk of M31 had been 
in place at early epoch based on the existence of metal-poor thin-disk 
GCs in the catalog of Perrett et al. (2002). However, Fusi Pecci et al. (2005) 
claim that there may be a systematic bias in  [Fe/H] determiation for 
young GCs by Perrett et al. (2002) due to the age-metallicity degeneracy 
effect. Fusi Pecci et al. (2005) suggest that these young GCs may be 
possibly more metal-rich by $\sim 1$ dex in [Fe/H]. Other hints of 
high metallicity for young thin-disk GCs supports this possibility   
(e.g., Williams \& Hodge 2001a; Beasley et al. 2004). If this is the case, 
the metallicity of young GCs are similar or slightly larger than that 
of old GCs. More spectroscopic observations 
of M31 young thin-disk GCs are required to clarify this issue.

\section{Intermediate-Age Globular Clusters}

Spectroscopic and spectrophotometric observations of M31 clusters 
support the existence of intermediate-age GCs with mean 
age $\sim$ 5 Gyr (Jiang et al. 2003; Burstein et al. 2004; 
Beasley et al. 2005; Puzia et al. 2005).  However, the age dating of 
extragalactic GCs via line index measurements is affected by 
(1) uncertainties in spectroscopic observations and (2) the degeneracy 
between age and HB morphology on the strength of Balmer lines 
(e.g., de Freitas Pacheco \& Barbuy 1995; Lee et al. 2000; 
Maraston et al. 2003; Thomas, Maraston, \& Bender 2003; Schiavon et al. 2004; 
Trager et al. 2005).  While stars near the MSTO region are the most 
dominant sources for the integrated strength of Balmer lines, blue HB 
stars also make a substantial contribution to the equivalent widths of 
these lines.  Enhancement of Balmer lines due to these He burning 
stars are maximized when the distribution of HB stars is centered at 
$(B-V)_0$ $\sim$ 0 or $T_{eff} \sim 9500$K.  The effect of old blue 
HB stars on the strong Balmer lines is such that the spectroscopic 
ages are significantly underestimated when such populations are 
present in GCs.  

In this respect, $FUV-V$ colors provide a useful tool for discriminating 
the intermediate-age subpopulation from old ($>$ 10 Gyr) and young ($<$ 1 Gyr) 
GCs within a wide range of metallicity. While old and young GCs show 
significantly large FUV to optical flux ratios, the intermediate-age GCs are 
relatively faint in FUV (see Fig. 2 of Lee, Lee, \& Gibson 2003). This is 
because intermediate-age GCs only contain warm MSTO stars which are not hot 
enough to produce a significant amount of FUV flux.  If these are truly 
intermediate-age objects as suggested by spectroscopic observations, then they 
should not be detected (or be very faint) in our {\sl GALEX} FUV photometry within 
the detection limit (Lee \& Worthey 2005).

Puzia et al. (2005) derived spectroscopic ages for 70 GCs in M31 based on 
Lick index measurements.  They find a population of intermediate-age GCs
with ages 5 -- 8 Gyrs and a mean metallicity of [Z/H] $\sim -0.6$. 
Independently, Burstein et al. (2004) and Beasley et al. (2005) 
also find two and six intermediate-age GCs in M31, repsectively.
Here, we consider 12  intermediate-age GCs with spectroscopically derived 
ages of 4 -- 7 Gyr from Table A.3 of Puzia et al. (2005).  There are four 
intermediate-age clusters in common between Beasley et al. (2005) and 
Puzia et al. (2005).  Most of the intermediate-age candidates have metallicities 
in the range [Fe/H] $< -1.0$ (Barmby et al. 2000).  Among the 16 intermediate-age
candidates identified by Puzia et al. (2005), Beasley et al. (2005), 
and Burstein et al. (2004), 12 and 10 GCs\footnote{Namely, they are 
B058, B110, B178, B182, B185 (NUV only), B126, B232, B292, B311, B337, 
B365, and NB67 (NUV only).  Four intermediate-age candidates (B301, NB16, NB81, 
and NB89) are not detected in both NUV and FUV.} are detected in $GALEX$
NUV and FUV passband, respectively. 

In Figure 14, we show $(B-V)_{0}$ vs. $(FUV-V)_{0}$ color-color diagram for 
the intermediate-age GC  candidates ({\it filled circles}) detected in FUV,
regardless of their E($B-V$) values. For comparison, we also plot old M31 
({\it open circles}) and Milky Way ({\it crosses}) GCs.  The model isochrones 
in Fig. 14 are for old (10, 12, and 14 Gyr; {\it dotted lines}), intermediate 
(5 Gyr; {\it solid line}), and young (1 Gyr; {\it long dashed line}) ages.  
The most striking feature of Fig. 14 is that all but one (B126) of the 
intermediate-age GC candidates detected in FUV are placed in the color-color diagram 
generally occupied by old GCs of age $\sim$ 12 -- 14 Gyr.  This suggests that most, 
if not all, of these intermediate-age GC candidates are in fact as old as the oldest M31 GCs.  
In this regard, it is important to note that intermediate-age cluster candidates 
B311 (Bursten et al. 2004) and B058 (Puzia et al. 2005) show clearly developed 
blue HBs in recent $HST$ CMDs by Rich et al. (2005). Individual CMDs for other 
intermediate-age GCs will give definitive answers to how the Balmer lines are 
affected by the presence of blue HB populations.

\section{Summary and Conclusions}

We present the $GALEX$ NUV and FUV photometry for GCs in M31 based on 
a mosaic of observations covering the whole of M31.
By cross-matching with objects in the RBC catalog, we detected 485 
and 273 GCs in NUV and FUV, respectively, which are the most complete 
and homogeneous UV data for M31 GCs.  We present our final UV catalog 
of M31 GCs, which is complemented with optical and near-infrared 
photometry for all UV-detected GCs.

We find that UV$-$optical color plays an important role as an age indicator 
while other possible contamination should be carefully considered 
(e.g., existence of extreme HB stars due to the helium variation among 
stellar populations in a GC; Lee et al. 2005 and references therein).  
At a given metallicity, the older populations with bluer HB stars are 
relatively UV bright, especially in the FUV passband. 
Comparing M31 data with Galactic GCs in the UV based on our population 
models that include the treatment of the detailed systematic variation 
of HB morphology with age and metallicity, we find that the age ranges 
of old GCs in M31 and Galactic halo are similar.  While deep HST imaging
of field populations in M31 find that disparate distant fields all show substantial
6-10 Gyr age ranges with only a minority of old stars (Brown et al. 2006) our
UV study of the GCS continues to confirm the view that the GCS of M31 is 
old and similar to that of the Milky Way, except for a minority of either 
very young or intermediate-age clusters. The {\sl GALEX} mission will rapidly 
enlarge the sample of GCs in nearby galaxies within 3 -- 4 Mpc and enable us 
to quantify any systematic mean age differences between cluster systems, 
modulo issues such as the effect of minority helium enriched populations.

Metal-rich GCs have a hint of wide dispersion of NUV colors, which 
reflects the variation of HB morphology and UV flux of metal-rich 
GCs in M31 due to parameters other than metallicity and age.
Furthermore, while most metal-rich GCs are faint in the FUV, we find
three candidate metal-rich ([Fe/H]$>-1$) GCs with significant FUV 
flux like that found for the peculiar Galactic GCs, NGC 6388 and NGC 6441, 
with extended blue HB stars. This indicates that even metal-rich M31 GCs 
may have hot blue HB populations rivaling any metal-poor GCs. Our discovery 
demonstrates for the first time that a possible population of hot extreme 
HB stars can exist in old, metal-rich GCs in M31.  The UV properties of 
UV bright metal-rich GCs in M31 and Milky Way are in good agreement with 
those of metal-rich GCs in M87, which suggests that these anomalous UV 
bright metal-rich GCs may be ubiquitous in many galaxies. 
In this regard, deep $HST$ ACS FUV observations (PID \#10901) for 
GCS of NGC 1399 will provide an opportunity to determine whether 
GCs in NGC 1399 also share characteristics with those in M87 
(and with UV bright metal-rich GCs in Milky Way and M31).
  
According to our population models, the strong UV flux of UV bright 
metal-rich GCs can be reproduced by even a small fraction of helium 
enhanced subpopulation in addition to the majority of the population 
having normal helium abundance within a GC.  We suggest that 
a reasonable agreement between observed UV colors of UV bright 
metal-rich GCs and population models can be achieved when we adopt 
a wide range of helium abundance. The discovery of UV bright 
metal-rich GCs is important in the context that the very hot extreme 
HB stars may be responsible for the upturn in FUV flux seen in luminous 
metal-rich elliptical galaxies and spiral bulges.

We confirm that the subsystem of blue GCs in M31 with thin-disk 
kinematics consists of dominantly young ($<$ 1 Gyr) GCs.  This is in 
line with the suggestion that significant formation of possible 
massive clusters was triggered recent epoch in the disk of M31. 
We also find a thin-disk subset among red GCs in M31 and their ages are 
likely old ($>$ 10 Gyr).  Based on our results from $GALEX$ UV 
observations and previous studies of M31 GCs (Fusi Pecci et al. 2005; 
Puzia et al. 2005), we suggest that the overall feature of M31 
thin-disk GCs regarding their formation and evolution is as 
follows:  a small fraction of GCs (i.e., red thin-disk GCs) belonging 
to the thin-disk are formed at an early epoch and 
maintain their kinematic properties. 
Significant formation of a large fraction of GCs (e.g. blue thin-disk 
GCs) appears to have occurred recently ($<$ 1 Gyr) in the 
thin-disk of M31. This leads us to conclude that the M31 and Milky Way
GCSs are even more different than had previously been suggested.

It is intriguing to note that the formation regions of young 
thin-disk GCs (e.g., Morrison et al. 2004) coincide with those of 
massive star formation in the outskirts of M31 observed by $GALEX$ 
(Thilker et al. 2005a) and $Spitzer$ (Gordon et al. 2006). 
This is also consistent with the most recent results of $GALEX$ 
observations that some nearby spiral galaxies (M33, M51, M83, and M101, 
Bianchi et al. 2005; Thilker et al. 2005a, b) show similar 
galactocentric gradients of $FUV-NUV$ color becoming bluer outward, 
which indicates younger stellar populations toward the outer parts 
of the galaxy disk.  While it may be tempting to 
consider a connection between young thin-disk GCs 
and the recently discovered extended disklike stellar structure 
around the M31 (Ibata et al. 2005), the age distribution of the M31
disk population at 30 kpc is still dominated by 4-10 Gyr old stars
(Brown et al. 2006).

The discovery of young GC-like objects in the disk of M31 provides 
an opportunity to study the ongoing process of GC formation at the 
current epoch and to connect with the properties of old GCs (Larsen 
2004).  While many observational results establish the presence of 
young massive clusters in a variety of environments that are peculiar 
in some ways (e.g., starburst galaxies and merger galaxies), the existence 
of young GCs in M31 disk may be evidence that young massive clusters 
can form even in the disks of normal spiral galaxies (see the recent review 
of Larsen 2004 and references therein).  In this respect, it is 
important to extend our UV studies of GCS to other galaxies such as 
M33, LMC/SMC, and Cen A.  Such studies are currently underway and 
the results will be presented in our forthcoming papers.

We detect 12 and 10 intermediate-age GC candidates classified by previous 
spectroscopic observations in NUV and FUV, respectively. 
On the basis of the UV photometry, we cautiously suggest that 
some of the spectroscopically identified-age GCs may not be truly 
intermediate-age objects, but rather older GCs with developed HB 
stars that contribute to enhanced UV flux as well as Balmer lines.  
We suggest that the FUV photometry can be an indirect indication
of the presence of intermediate-age GCs and may provide a complementary test of 
the spectroscopic observations for the age dating of GCs.  Since 
spectroscopic results and UV photometry are based on the integrated 
properties of stellar population, deep $HST$ imaging with high 
spatial resolution for intermediate-age GC candidates is critical in order to 
derive their HB morphologies.  In this respect, the outcomes of 
the $HST$ ACS observations (PID \#10631) of intermediate-age GC candidates 
(B337, B336, B058, B049, B206D, B292, and B350) of 
M31 are highly anticipated.  This will provide a definite answer to 
the question of whether spectroscopically derived ages for intermediate-age GCs are 
due to the truly younger ages or the anomalous hot HB stars.  
In any case, our $GALEX$ UV photometry of the intermediate-age 
GC  candidates will 
have important implications for spectroscopically derived age and 
metallicities of clusters in distant stellar populations.

\acknowledgments
We are grateful for the clarifications and improvements
suggested by an anonymous referee.
GALEX (Galaxy Evolution Explorer) is a NASA Small Explorer, launched in 
April 2003. We gratefully acknowledge NASA's support for construction, 
operation, and science analysis for the GALEX mission, developed in 
cooperation with the Centre National d'Etudes Spatiales of France and 
the Korean Ministry of Science and Technology. Yonsei University participation 
was supported by the Creative Research Initiative Program of MOST/KOSEF, 
for which we are grateful. This work was supported by the Korea Research 
Foundation Grant funded by the Korean Government (MOEHRD) (KRF-2005-202-C00158).

%{\it Facilities:} \facility{GALEX}

\clearpage

%%% FIGURES

% Fig. 1
\begin{figure}
\epsscale{0.5}
\plotone{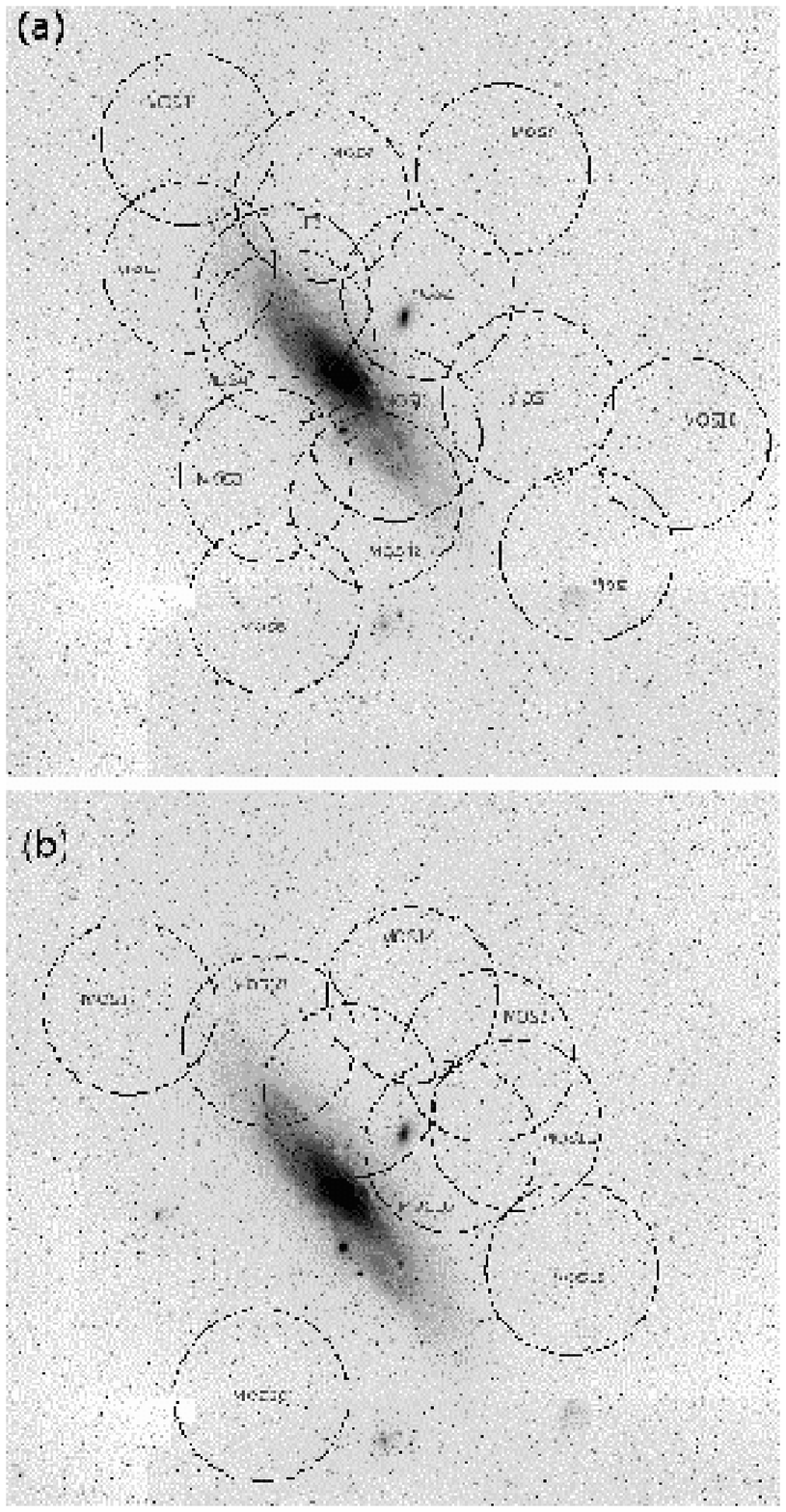}
\caption{ (a) 14 $GALEX$ fields observed in 2003 and (b) additional 9 $GALEX$ fields
          observed in 2004 projected onto a 6\arcdeg$\times$6\arcdeg DSS image. 
          North is to the top and East is to the left.  
          }
\label{f:f1}
\end{figure}

%Fig. 2
\begin{figure}
\epsscale{0.6}
\plotone{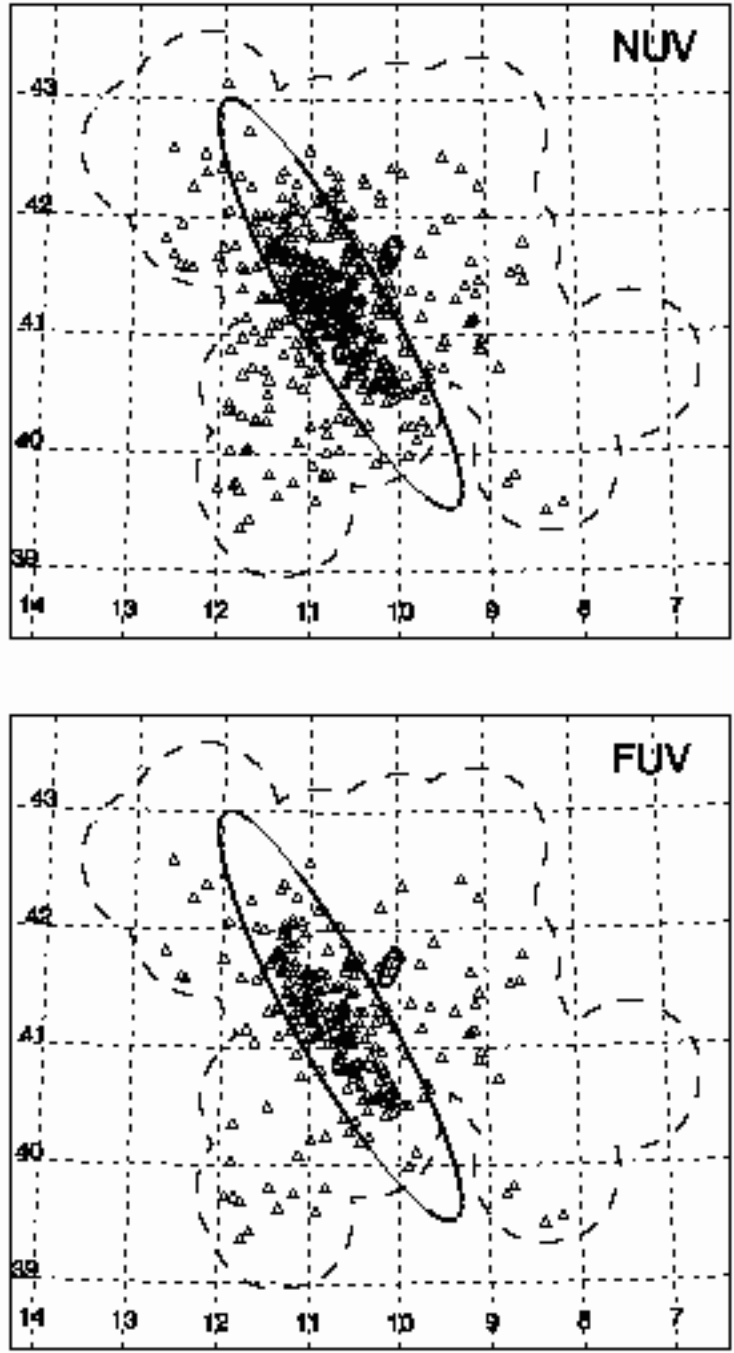}
\caption{ Spatial distribution on the sky of objects detected in $GALEX$ NUV and FUV fields.
          The large ellipse in solid line is the disk/halo criterion for M31 defined
          by Racine (1991); two smaller ellipses are the $D_{25}$ isophotes of NGC 205 ($larger$)
          and M32 ($smaller$).  The boundary in dashed line shows the entire field covered by
          $GALEX$ observations.
          }
\label{f:f2}
\end{figure}

%Fig. 3
\begin{figure}
\epsscale{0.6}
\plotone{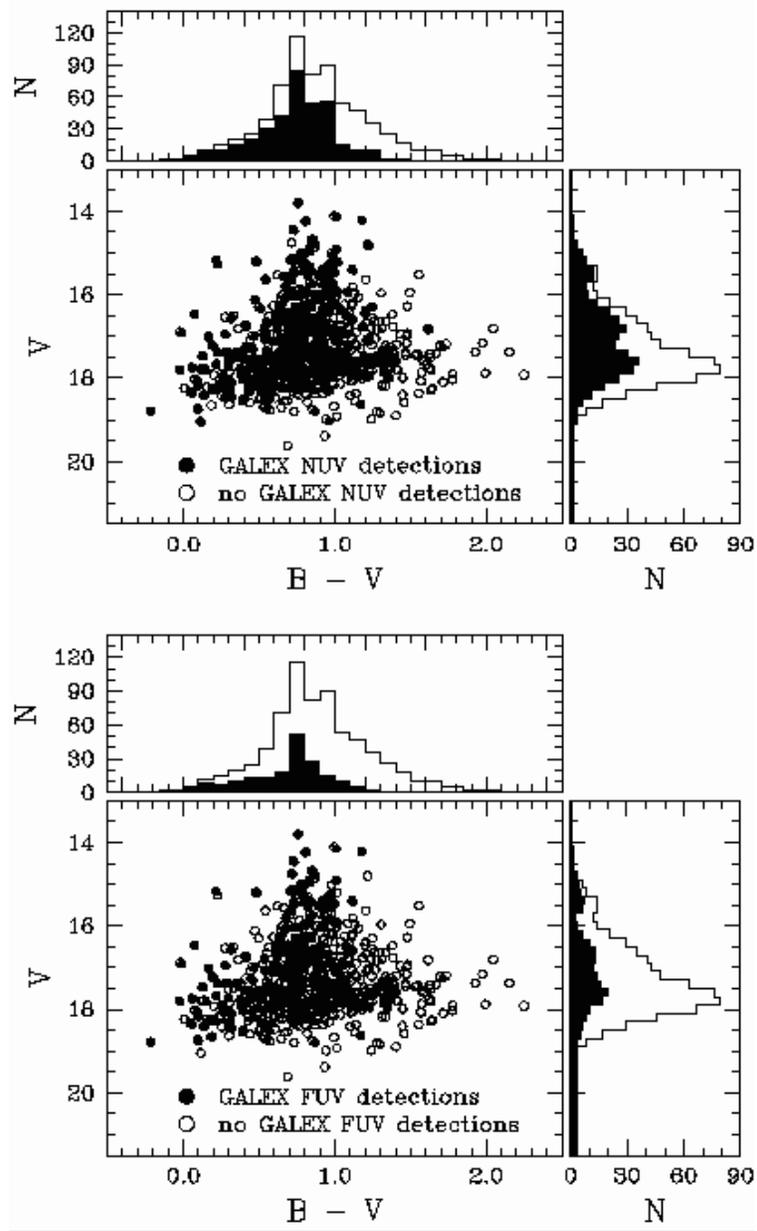}
\caption{ $GALEX$ detection rates in NUV ({\it upper}) and FUV ({\it lower}).
          Of the 693 objects with both $B$ and $V$ data in RBC, 485 (about 70\%) and 273 (about 39\%) 
          objects have their entries in the $GALEX$ UV catalog for NUV and FUV, respectively.
          }
\label{f:f3}
\end{figure}

%Fig. 4
\begin{figure}
\epsscale{1.0}
\plotone{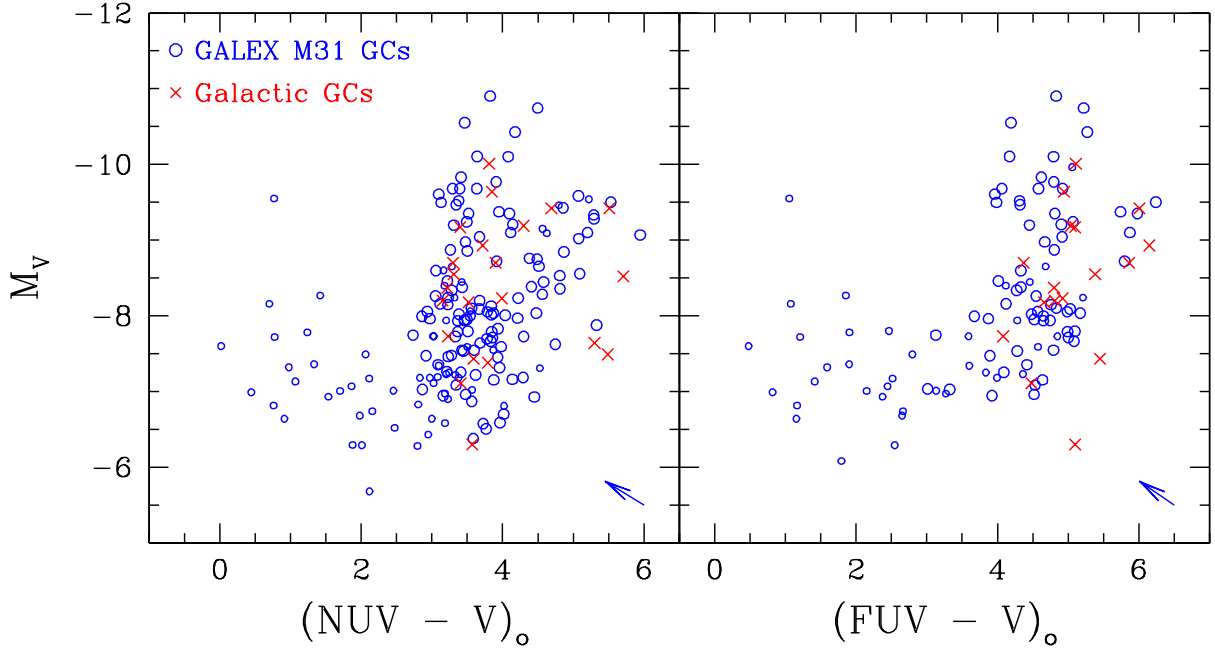}
 \caption{
M$_{V}$ vs. $(UV-V)_0$  color-magnitude diagrams for {\sl GALEX} M31 GCs ({\it circles}).
Only confirmed M31 GCs in RBC entries (i.e., class 1) are selected.
Large circles are GCs with E($B-V$) $<$ 0.16 from Barmby et al. (2000).
Small circles are GCs with no available reddening information in Barmby et al., 
assuming that they are only affected by the foreground Galactic reddening of 
E($B-V$)=0.10. {\sl GALEX} M31 GCs are compared with Galactic GCs 
obtained from OAO-2 and ANS ({\it crosses}; Sohn et al. 2006).  
Relatively red GCs [$(NUV-V)_0$ $\geq$ 3.0 and $(FUV-V)_0$ $\geq$ 4.0] in M31  
and old Galactic GCs occupy the same area in CMDs. Blue clusters 
[$(NUV-V)_0$ $\leq$ 3.0 and $(FUV-V)_0$ $\leq$ 4.0] are young cluster candidates 
(see text). The arrow indicates reddening vector by an increase of E($B-V$) = 0.10.
}
\label{f:f4}
\end{figure}

%Fig. 5 
\clearpage
\begin{figure}
\epsscale{1.0}
\plotone{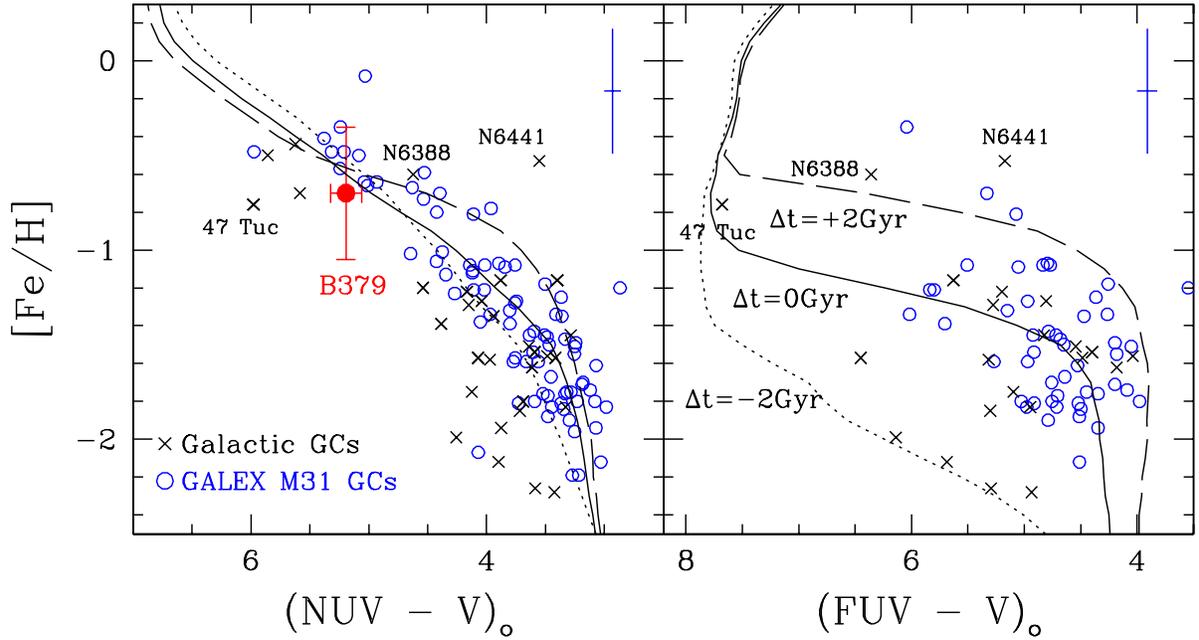}
 \caption{
[Fe/H] vs. $(UV-V)_{0}$ diagrams for M31 ({\it open circles}) and 
Galactic GCs ({\it crosses}; [Fe/H] from Harris 
1996).  The model isochrones are constructed from our evolutionary 
population models of GCs in the $GALEX$ filter system 
(Chung, Yoon, \& Lee 2006, in preparation; see also Lee et 
al. 2002; Yi 2003).  The $\Delta t$ = 0 ({\it solid line}) isochrone 
corresponds to inner halo Galactic GCs (Galactocentric radius 
$\leq$ 8 kpc) of $\sim 12$ Gyr.  The $\Delta t$ = +2 Gyr ({\it 
long dashed line}) and $-2$ Gyr ({\it dotted line}) isochrones 
are for the models 2 Gyr older and younger than the inner halo 
Galactic GCs, respectively.  Typical errors of [Fe/H] 
[calculated from column (20) of Table 2] and colors of M31 GCs are 
shown as error bars.  Cluster B379, an M31 GC with 
direct age estimation by Brown et al. (2004), is shown as a 
filled circle.}
\label{f:f5}
\end{figure}

% Fig. 6
\begin{figure}
\epsscale{1.0}
\plotone{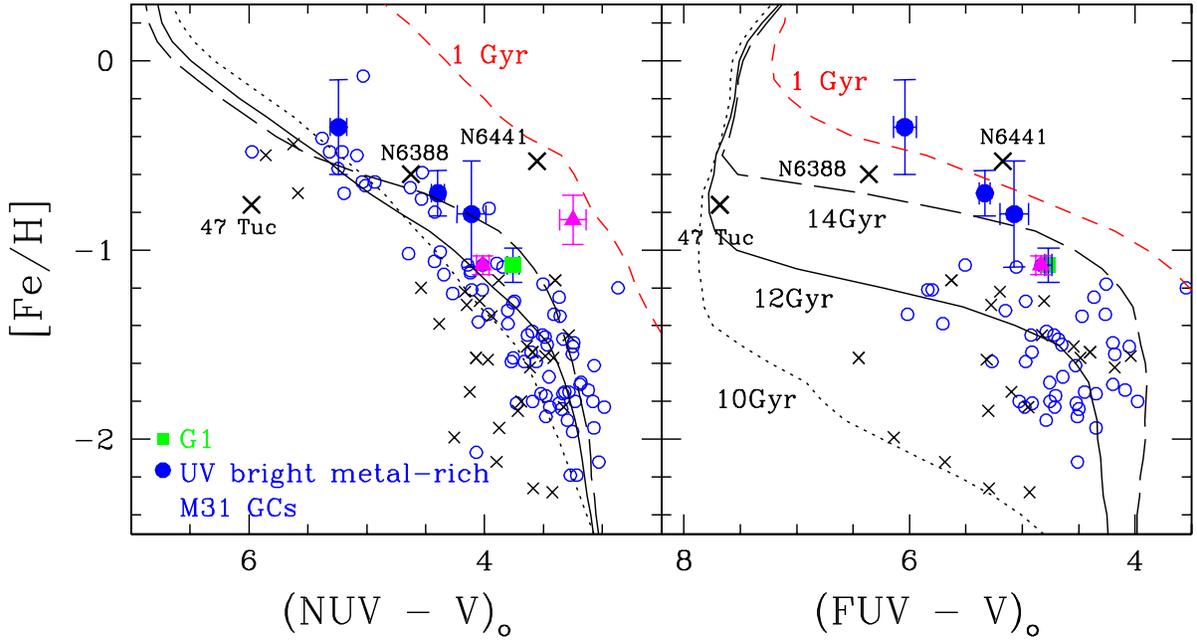}
 \caption{
[Fe/H] vs. $(UV-V)_0$ diagrams for UV bright metal-rich GCs in M31 ({\it filled circles}).
Three metal-rich ([Fe/H] $> -1.0$) M31 GCs show significant FUV flux comparable to those 
of peculiar metal-rich Galactic GCs, NGC 6388 and NGC 6441, with anomalous hot HB stars.
The filled triangles indicates metal-rich M31 GCs (B158 and B234) that may be 
analogs of NGC 6388 and NGC 6441 suggested by Beasley et al. (2005) from the results of 
Ca II index. The moderately metal-rich ([Fe/H] = $-1.08$) M31 GC G1 ({\it filled square}) 
shows relatively large UV flux which is consistent with its HB morphology with a notable minor 
population of blue HB stars. We also superpose a young (1 Gyr, {\it dashed line}) isochrone.
}
\label{f:f6}
\end{figure}

%Fig. 7
\begin{figure}
\epsscale{1.0}
\plotone{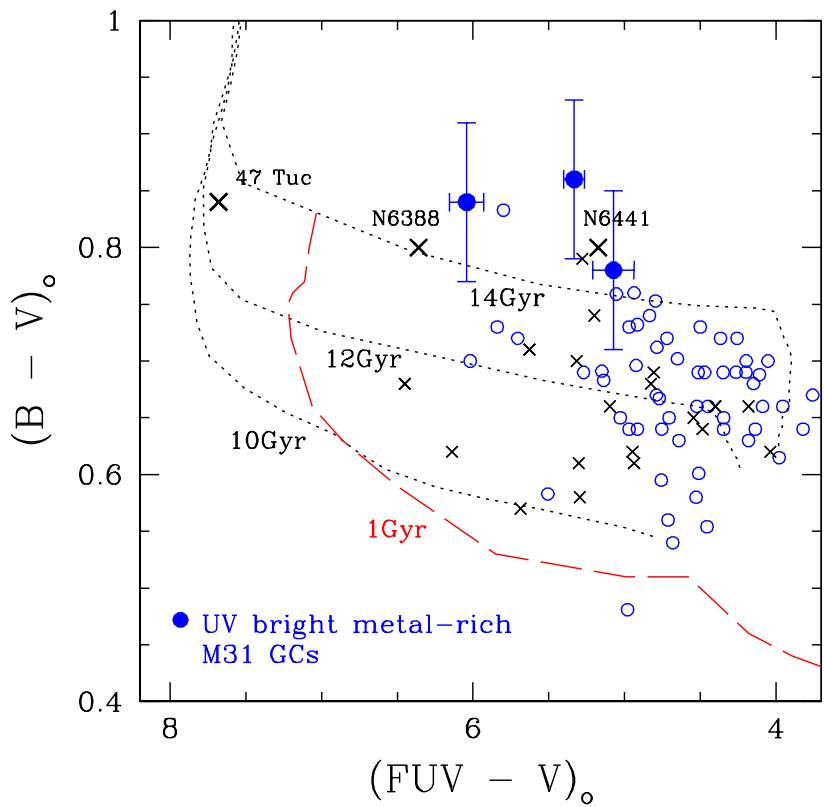}
 \caption{
$(B-V)_0$ vs. $(FUV-V)_0$ diagram for UV bright metal-rich GCs in M31 ({\it filled circles}).
The long dashed line is for the 1 Gyr isochrone. All candidate UV bright metal-rich GCs 
follow isochrones for old ages, which confirms that they are bona-fide old metal-rich GCs with 
large FUV flux. The $B-V$ and $FUV-V$ errors of the UV bright metal-rich GCs are estimated from 
the typical internal errors (0.05 mag) of $B$ and $V$ magnitudes suggested by Galleti et al. (2004).
}
\label{f:f7}
\end{figure}

%Fig. 8
\clearpage

\begin{figure}
\epsscale{1.0}
\plotone{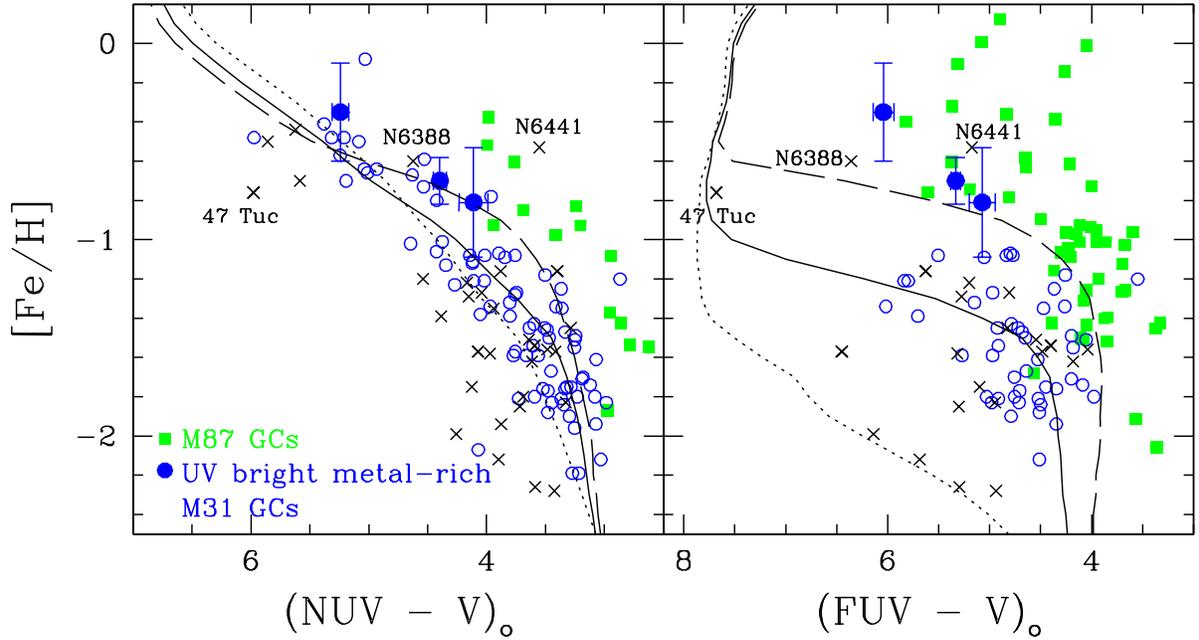}
 \caption{
Comparison of UV bright metal-rich GCs in M31 ({\it filled circles}) with GCs in giant elliptical 
galaxy M87 ({\it filled squares}; Sohn et al. 2006). Most M87 GCs are systematically UV bright at a 
fixed [Fe/H], compared to the Milky Way ({\it crosses}) and M31 ({\it circles}) GCs. It is worth 
to note that metal-rich UV bright GCs in M31 and Galactic counterparts, NGC 6388 and NGC 6441, 
fall in the vicinity of the distribution of metal-rich M87 GCs in [Fe/H] vs. $(FUV-V)_0$ diagram. 
This suggests that UV bright metal-rich GCs in M31 and Milky Way may share their properties with 
metal-rich GCs in M87.
}
\label{f:f8}
\end{figure}

%Fig. 9
\begin{figure}
\epsscale{1.0}
\plotone{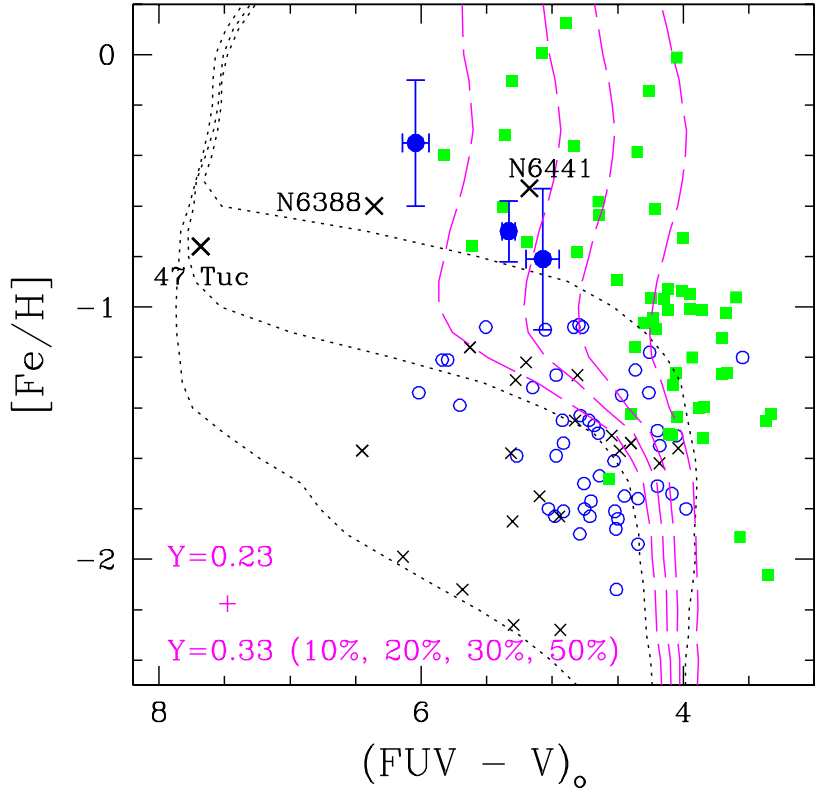}
 \caption{
Comparison of UV bright metal-rich  GCs in M31 ({\it filled circles}) and M87 ({\it filled squares}; 
Sohn et al. 2006) with the model predictions. Dotted lines are for the case in which all the GCs 
have same primodial helium abundance of Y = 0.23 but different ages. Dashed lines are for the case with 
minority population of helium enhancement (Y = 0.33) in addition to the majority of the population 
with normal helium abundance (Y = 0.23) within a model GC at the same age of 12 Gyr. 
Different dashed lines indicate GCs with different subpopulation number fraction of helium enhancement 
(left to right, 10\%, 20\%, 30\%, and 50\%) within a GC. A reasonable agreement between observed 
UV colors of metal-rich GCs and models is obtained when we adopt a large range of 
helium abundance.
}
\label{f:f9}
\end{figure}

%Fig. 10

\begin{figure}
\epsscale{1.0}
\plotone{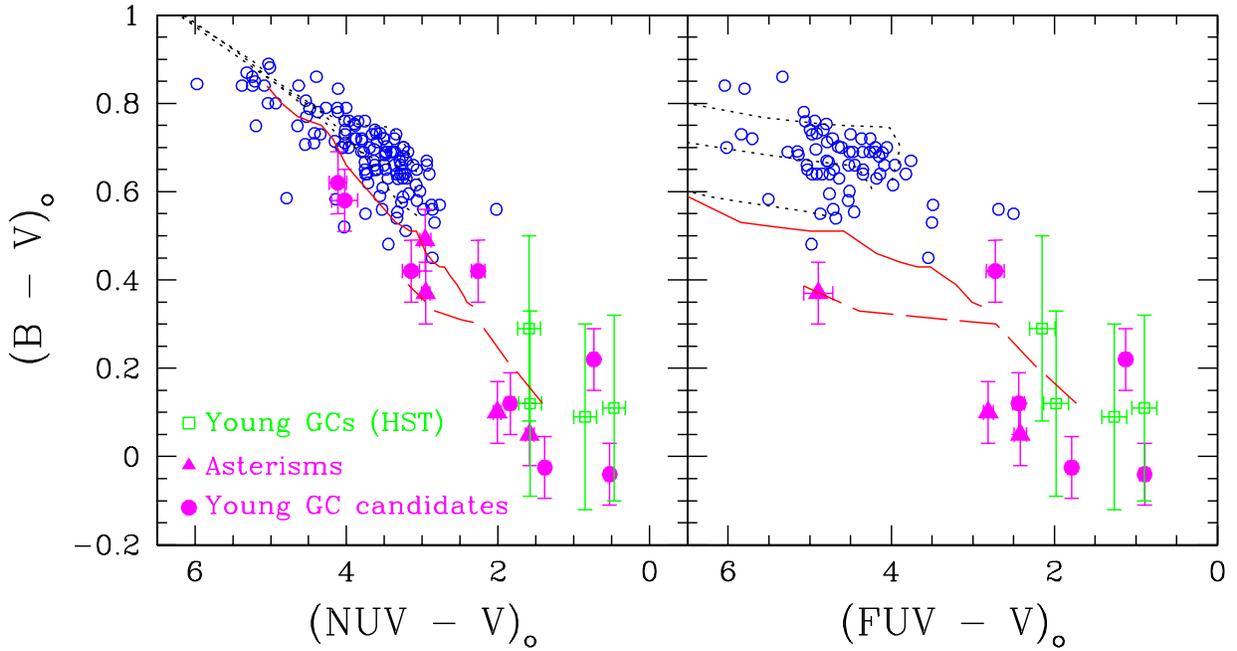}
 \caption{
$(B-V)_0$ vs. $(UV-V)_0$ diagrams for young GCs. Filled circles are young GCs found from 
the recent high-quality spectroscopic observations (Beasley et al. 2004; Burstein et al. 2004).
We adopt E($B-V$) = 0.10 for these clusters. The spectroscopically determined ages of young GCs 
are in good agreement with our population models in the range of 0.5 Gyr ({\it long dashed line}) 
-- 1 Gyr ({\it solid line}). Open squares are young GCs observed by Williams \& Hodge (2001a) 
with {\sl HST}. Filled triangles are probable asterisms suggested by Cohen et al. (2005). 
Open circles are possible old GCs in M31 and dotted lines are our model isochrones in the range
of 10 Gyr -- 14 Gyr.
}
\label{f:f10}
\end{figure}

% Fig. 11
\clearpage

\begin{figure}
\epsscale{1.0}
\plotone{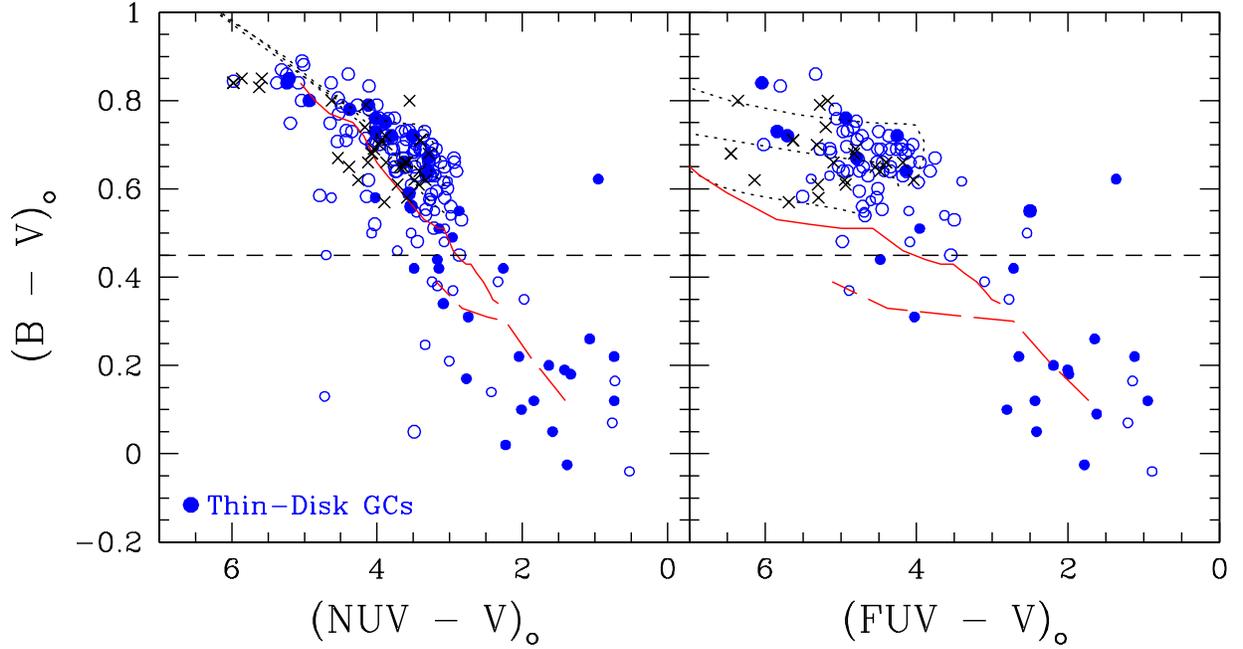}
 \caption{
$(B-V)_0$ vs. $(UV-V)_0$ diagrams for M31 GCs with thin-disk kinematics ({\it filled circles})
from Morrison et al. (2004). We superpose our model isochrones of 1 Gyr ({\it solid line}) and 0.5 Gyr 
({\it long dashed line}). Dotted lines are our model isochrones in the range of 10 Gyr -- 14 Gyr.
The dashed horizontal line corresponds to the reference value of $(B-V)_0$ = 0.45 for young GC 
selection adopted from Fusi Pecci et al. (2005). Large fraction of thin-disk GCs with $(B-V)_0 <$ 0.45 
show young ages with less than 1 Gyr. At the same time, small fraction of thin-disk GCs with 
$(B-V)_0 >$ 0.45 appear to be old. 
}
\label{f:f11}
\end{figure}

% Fig. 12
\begin{figure}
\epsscale{1.0}
\plotone{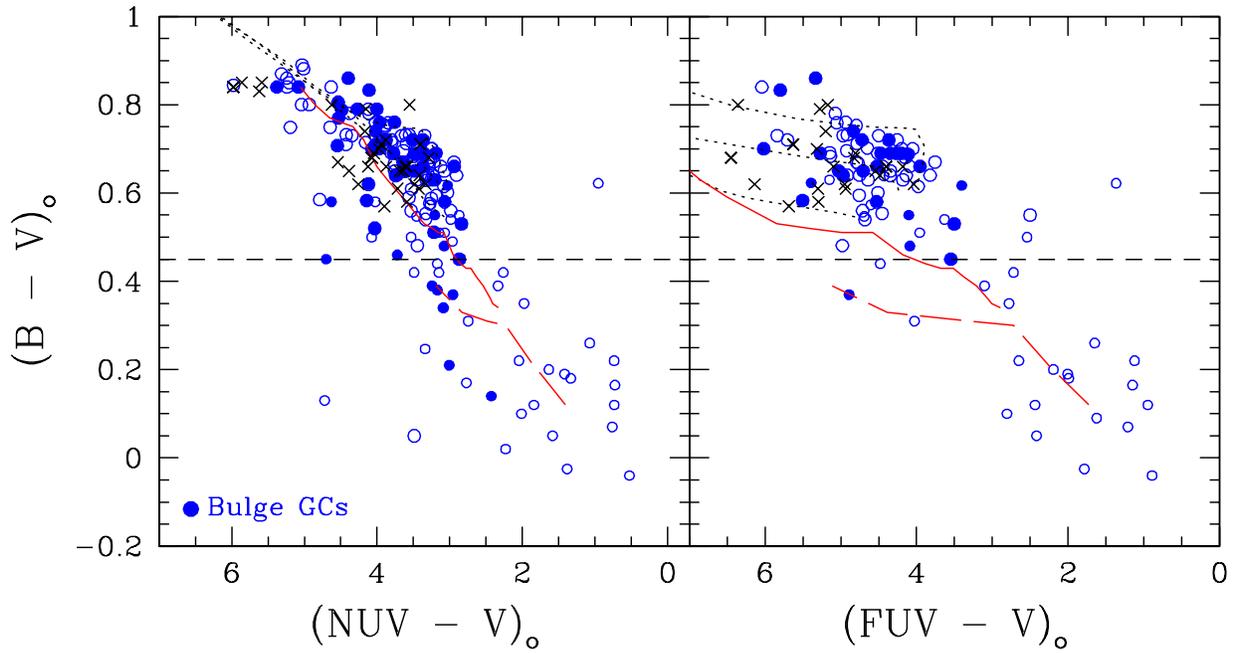}
 \caption{
Same as Fig. 11, but for M31 GCs with bulge kinematics ({\it filled circles}) from Morrison et al. (2004). 
Contrary to the thin-disk GCs, bulge GCs are biased to the relatively red optical/UV colors 
which indicates the mean age of bulge GCs is old. The age distribution of bulge GCs support most of blue GCs 
[$(B-V)_0 <$ 0.45] without kniematic data are possible thin-disk clusters. 
}
\label{f:f12}
\end{figure}

% Fig. 13

\begin{figure}
\epsscale{1.0}
\plotone{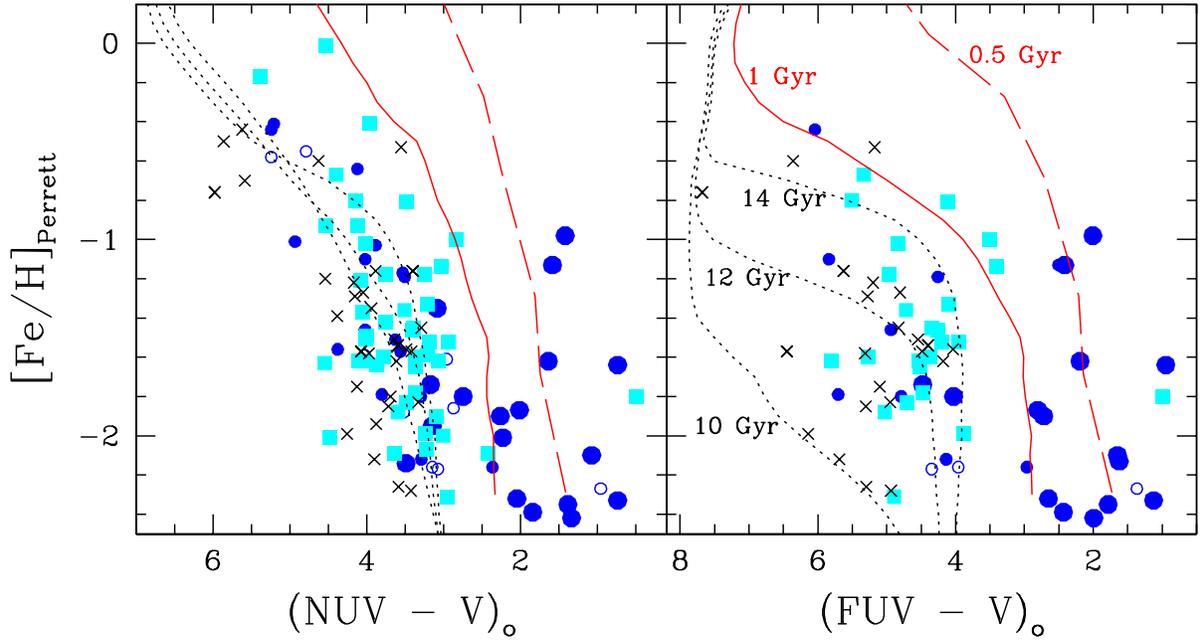}
 \caption{
[Fe/H] vs. $(UV-V)_0$  diagrams for thin-disk GCs ({\it filled circles}) and bulge GCs ({\it filled squares}) 
in M31. In addition to the model isochrones with old ages comparable to mean Galactic halo 
(10 -- 14 Gyr, {\it dotted lines}), we also superpose our young isochrones of 1 Gyr ({\it solid line}) and 
0.5 Gyr ({\it long dashed line}). Optically blue thin-disk GCs [$(B-V)_0 <$ 0.45; {\it large filled circles}] 
show systematically stronger UV flux than do red thin-disk GCs [$(B-V)_0 >$ 0.45; {\it small filled circles}] 
at a given metallicity. This is consistent with that the blue thin-disk GCs are systematically young GCs 
with $<$ 1 Gyr. GCs with bulge kinematics are biased to the red UV color distribution, which is an indication 
of their old ages. Since [Fe/H] values of Barmby et al. are not available for many thin-disk GCs of 
Morrison et al. (2004), we instead adopt [Fe/H] from Perrett et al (2002).
}
\label{f:f13}
\end{figure}

% Fig. 14
\begin{figure}
\epsscale{1.0}
\plotone{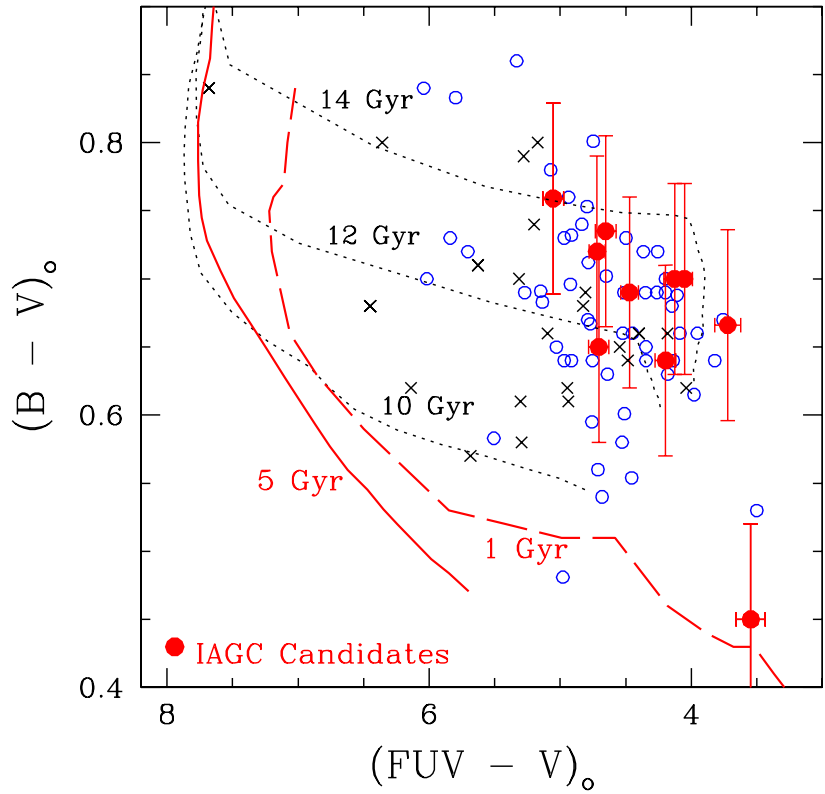}
 \caption{
Comparison of spectroscopically classified intermediate-age GC candidates ({\it filled circles}) 
with the model predictions. Superposed are our model isochrones for old (10, 12, and 14 Gyr; {\it dotted lines}), 
intermediate (5 Gyr; {\it solid line}), and young (1 Gyr; {\it long dashed line}) ages. All but one 
intermediate-age GC candidates detected in FUV are placed in the color-color diagram 
generally occupied by old GCs of age with 12 -- 14 Gyr. 
}
\label{f:f14}
\end{figure}

\clearpage

\begin{deluxetable}{lccc}

%\tablenum(1)
\setcounter {table}{0}
\tablewidth{0pt}
\tablecaption{Position of {\sl GALEX} M31 Additional Fields}
\tablehead{
\colhead{Field}       &      \colhead {RA (2000)}  &
\colhead {DEC (2000)}   &    \colhead {Exposure (s)}}
%\scriptsize
\startdata
     M31-MOS13  &  00 36 00.00  &  41 45 00.00   &  1620  \\
     M31-MOS14  &  00 40 00.00  &  42 42 00.00   &  2338  \\
     M31-MOS15  &  00 33 55.20  &  40 42 00.00   &  3241  \\
     M31-MOS16  &  00 38 31.20  &  41 36 00.00   &  5113  \\
     M31-MOS17  &  00 42 24.00  &  42 00 00.00   &  5108  \\
     M31-MOS18  &  00 45 36.00  &  42 21 36.00   &  3251  \\
     M31-MOS19  &  00 51 07.20  &  42 34 12.00   &  3231  \\
     M31-MOS20  &  00 45 40.80  &  39 47 06.00   &  1456  \\
     M31-MOS21  &  00 37 01.20  &  42 15 00.00   &  1357
\enddata
\end{deluxetable}

\clearpage
%\LongTables
%\begin{landscape}
\thispagestyle{empty}
\begin{deluxetable}{lllcccccccccccccccccc}
\setlength{\tabcolsep}{0.11cm}
%\tabletypesize{\tiny}
%\rotate
\tablecaption{ UV and optical photometry of M31 globular clusters and candidates\label{t:table2}}
\tablehead{ \colhead{}   & \colhead{}        &    \colhead{}      & \colhead{}      &   \colhead{RA}      & \colhead{DEC}     & \colhead{} 
& \colhead{}                     & \colhead{}                & \colhead{}                     & \colhead{}    & \colhead{}    & \colhead{}  
& \colhead{}    & \colhead{}    & \colhead{}    & \colhead{}    & \colhead{}    & \colhead{}       & \colhead{}  & \colhead{Residual} \\
            \colhead{ID} & \colhead{Alt. ID}  & \colhead{NED ID} & \colhead{Flag} & \colhead{(J2000)} & \colhead{(J2000)} & \colhead{$m_{\tiny NUV}$} & \colhead{$\sigma_{\tiny NUV}$} & \colhead{$m_{\tiny FUV}$} & \colhead{$\sigma_{\tiny FUV}$} & \colhead{$U$} & \colhead{$B$} & \colhead{$V$} & \colhead{$R$} & \colhead{$I$} & \colhead{$J$} & \colhead{$H$} & \colhead{$K$} & \colhead{[Fe/H]} & \colhead{$\sigma_{\rm [Fe/H]}$} & \colhead{(km s${-1}$)}
          }
\startdata
G001    & 000-001    &   SKHV 001    &   1   & 00:32:46.51   & +39:34:39.7 &	18.015 &     0.012 &	18.972 &     0.031    &   \nodata &	14.56 &     13.81 &	13.21 &     12.68 &	11.85 &     11.13 &	11.04 &     -1.08 &	 0.09 &   \nodata \\
G002    & 000-002    &   SKHV 002    &   1   & 00:33:33.79   & +39:31:18.5 &	19.383 &     0.023 &	20.911 &     0.079    &     16.75 &	16.48 &     15.81 &	15.32 &   \nodata &	14.26 &     13.71 &	13.58 &     -1.70 &	 0.36 &   \nodata \\
B289    & \nodata    &   Bol 289     &   1   & 00:34:20.85   & +41:47:50.9 &	19.861 &     0.030 &	21.140 &     0.092    &     17.00 &	16.76 &     16.09 &	15.61 &     14.97 &	14.52 &     13.96 &	13.96 &   \nodata &   \nodata &   \nodata \\
B290    & 290-000    &   Bol 290     &   1   & 00:34:20.86   & +41:28:17.9 &	22.500 &     0.116 &   \nodata &   \nodata    &     18.46 &	18.04 &     17.14 &	16.57 &     15.72 &	15.30 &     14.48 &	14.33 &   \nodata &   \nodata &   \nodata \\
B411    & \nodata    &   Bol 411     &   2   & 00:34:30.83   & +41:33:43.8 &	20.753 &     0.035 &	21.361 &     0.074    &     18.59 &	18.81 &     17.75 &	17.42 &     16.42 &	15.85 &     14.73 &	14.42 &   \nodata &   \nodata &   \nodata \\
BA21    & \nodata    &   [BA64] 3-21 &   2   & 00:34:54.01   & +39:49:41.5 &	20.954 &     0.054 &	21.774 &     0.124    &     17.56 &	17.53 &     16.64 &   \nodata &   \nodata &	15.09 &     14.37 &	13.87 &   \nodata &   \nodata &   \nodata \\
B412    & \nodata    &   Bol 412     &   2   & 00:34:55.22   & +41:32:25.5 &	21.754 &     0.067 &	22.299 &     0.190    &     18.93 &	18.43 &     17.36 &	16.74 &     15.86 &	14.93 &     14.11 &	13.95 &   \nodata &   \nodata &   \nodata \\
BA22    & \nodata    &   [BA64] 3-22 &   2   & 00:35:13.77   & +39:45:38.3 &	21.467 &     0.074 &	21.840 &     0.131    &   \nodata &   \nodata &   \nodata &   \nodata &   \nodata &	15.99 &     15.12 &	14.96 &   \nodata &   \nodata &   \nodata \\
B134D   & \nodata    &   Bol D134    &   2   & 00:35:30.33   & +40:44:26.1 &	20.785 &     0.028 &	21.608 &     0.069    &     18.46 &	18.65 &     18.19 &	17.40 &   \nodata &	15.61 &     15.34 &	14.60 &   \nodata &   \nodata &   \nodata \\
B291    & 291-009    &   Bol 291     &   1   & 00:36:04.87   & +42:02:09.8 &	20.916 &     0.039 &   \nodata &   \nodata    &     17.65 &	17.34 &     16.59 &	16.08 &     15.38 &	14.70 &     14.04 &	14.13 &   \nodata &   \nodata &   \nodata \\
B292    & 292-010    &   Bol 292     &   1   & 00:36:16.59   & +40:58:26.6 &	21.018 &     0.033 &	21.796 &     0.085    &     17.87 &	17.89 &     17.00 &	16.62 &     16.06 &	15.41 &     15.02 &	14.73 &     -1.42 &	 0.16 &   \nodata \\
B414    & \nodata    &   Bol 414     &   2   & 00:36:19.31   & +42:16:30.6 &	20.611 &     0.023 &	20.998 &     0.043    &     18.56 &	18.48 &     17.98 &   \nodata &     16.95 &   \nodata &   \nodata &   \nodata &   \nodata &   \nodata &   \nodata \\
B293    & 293-011    &   Bol 293     &   1   & 00:36:20.78   & +40:53:36.6 &	20.141 &     0.021 &	21.174 &     0.057    &     17.19 &	17.03 &     16.30 &	15.79 &     15.24 &	14.71 &     14.12 &	14.15 &     -1.89 &	 0.17 &   \nodata \\
B137D   & \nodata    &   Bol D137    &   2   & 00:36:20.90   & +40:55:59.6 &	22.272 &     0.082 &   \nodata &   \nodata    &     19.00 &	19.05 &     18.51 &	18.06 &   \nodata &   \nodata &   \nodata &   \nodata &   \nodata &   \nodata &   \nodata \\
B138D   & \nodata    &   Bol D138    &   2   & 00:36:21.67   & +41:28:32.6 &	22.380 &     0.064 &	23.110 &     0.176    &     18.70 &	17.92 &     16.87 &	16.14 &   \nodata &	14.22 &     13.47 &	13.20 &   \nodata &   \nodata &   \nodata \\
\enddata
\tablenotetext{*} {The complete version of this table is in the electronic edition of the Journal.
The printed edition contains only a sample.}
\end{deluxetable}
\clearpage
%\end{landscape}

\end{document}